\documentclass[a4paper,11pt]{article}
\usepackage{jheppub} 
\usepackage{lineno}
\linenumbers

\usepackage{bm}
\usepackage{braket}
\usepackage[subrefformat=parens]{subcaption}

\newcommand{\coloneqq}{:=}
\newcommand{\no}{\nonumber}
\newcommand{\dm}{\mathrm{d}}

\title{Stochastic tunneling in de Sitter spacetime}

\author[\natural]{Taiga Miyachi,}
\author[\natural,\dagger]{Jiro Soda}
\author[\flat]{and Junsei Tokuda}
\affiliation[\natural]{Department of Physics, Kobe University,\\ 
Kobe 657-8501, Japan}
\affiliation[\dagger]{International Center for Quantum-field Measurement Systems for Studies of the Universe and Particles (QUP),\\ 
KEK, Tsukuba 305-0801, Japan }
\affiliation[\flat]{Particle Theory and Cosmology Group, Center for Theoretical Physics of the Universe,\\ 
Institute for Basic Science (IBS), Daejeon, 34126, Korea}

\emailAdd{tmiyachi@stu.kobe-u.ac.jp}
\emailAdd{jiro@phys.sci.kobe-u.ac.jp}
\emailAdd{jtokuda@ibs.re.kr}

\abstract{Tunneling processes in de Sitter spacetime are studied by using the stochastic approach. We evaluate the the Martin–Siggia–Rose–Janssen–de Dominicis (MSRJD) functional integral by using the saddle-point approximation to obtain the tunneling rate.
The applicability conditions of this method are clarified using the Schwinger-Keldysh formalism. In the case of a shallow potential barrier, we reproduce the Hawking-Moss (HM) tunneling rate. Remarkably, in contrast to the HM picture, the configuration derived from the MSRJD functional integral satisfies physically natural boundary conditions. We also discuss the case of a steep potential barrier and find an interesting Coleman-de Luccia (CDL) bubble-like configuration. 
Since the starting point of our analysis is the Schwinger-Keldysh path integral, which can be formulated in a more generic setup and incorporates quantum effects, our formalism sheds light on further studies of tunneling phenomena from a real-time perspective.
}

\makeatletter
\gdef\@fpheader{}
\makeatother

\begin{document}
\nolinenumbers 

\unitlength = 1mm
\begin{flushright}
KOBE-COSMO-23-09
\, CTPU-PTC-23-33
\end{flushright}

\maketitle
\flushbottom

\section{Introduction}

Tunneling is an important non-perturbative phenomenon
in cosmology. Hence, it is crucial to understand tunneling processes in curved spacetime. In particular, tunneling in de Sitter spacetime is worth studying in detail. 
The reason is that de Sitter spacetime is the simplest non-trivial curved spacetime. Moreover, the results obtained there can be applicable to tunneling phenomena during inflation. 
Since both Minkowski and de Sitter spacetime have the maximal symmetry, there exist a similarity.
Indeed, the Euclidean instanton method for evaluating the tunneling rate in Minkowski spacetime~\cite{Coleman:1977py,Callan:1977pt}
 is applicable to tunneling processes in de Sitter spacetime~\cite{Coleman:1980aw}.
However, the extension of Euclidean method to de Sitter spacetime has a difficulty in interpretation.
In the case of de Sitter spacetime, there are two saddle solutions: Coleman-De Luccia (CDL) \cite{Coleman:1980aw} and Hawking-Moss (HM) instantons \cite{Hawking:1981fz}.
The latter instanton is a homogeneous solution.
Therefore, it is not straightforward to interpret the tunneling rate as that for the tunneling process from a false vacuum to a true vacuum~\cite{Rubakov:1999ir,Brown:2007sd}.

To circumvent the difficulty,  a real time formalism might be useful.
In the case of de Sitter spacetime in the flat chart, 
the stochastic formalism  has been used
for investigating time evolution of fluctuations during inflation~\cite{Starobinsky:1986fx,Starobinsky:1994bd}.
In this formalism, Langevin equations describe the dynamics of long wavelength modes. Stochastic noises stem from short wavelength quantum modes. 
The stochastic formalism has been also used to analyze the tunneling processes in de Sitter spacetime~\cite{Goncharov:1987ir,Linde:1991sk,Tolley:2008qv,Noorbala:2018zlv,Hashiba:2020rsi,Camargo-Molina:2022ord,Camargo-Molina:2022paw}.
However, most of the previous works  treated only the HM transition by utilizing the Fokker-Plank (FP) equation for a homogeneous scalar field, which can be derived from the Langevin equation.
Obviously, the HM transition cannot describe the bubble nucleation in contrast to CDL tunneling. 
To describe the bubble, we have to include the spatial gradient term and the FP  equation becomes the functional differential equation instead of the partial differential equation.
Hence, at first glance, it seems difficult to treat the CDL bubble by the stochastic approach.
However, we make an observation that there exists a path integral representation to solutions of the FP equation. In condensed matter physics, it is called
 Martin–Siggia–Rose–Janssen–de Dominicis (MSRJD) functional integral~\cite{Martin:1973zz,DeDominicis:1975gjb,Janssen:1976aa,DeDominicis:1977fw}. 

\begin{figure}[btp]
\centering
\includegraphics{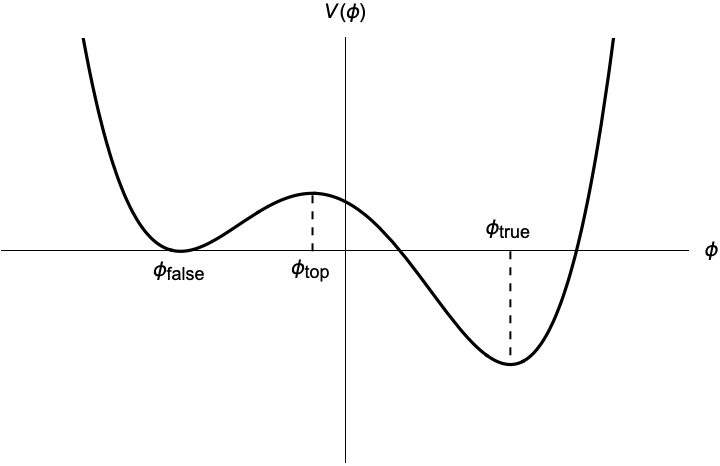}
\caption{The schematic picture of the potential $V(\phi)$ which has a false vacuum at $\phi=\phi_{\text{false}}$, true vacuum at $\phi=\phi_{\text{true}}$ and the top of the potential at $\phi=\phi_{\text{top}}$. \label{fig:potential_schematic}}
\end{figure}

In this paper, we use the MSRJD path integral formula to study the tunneling processes on the fixed de Sitter background. The applicability conditions of this method are clarified based on the Schwinger-Keldysh formalism. 
We consider a potential $V(\phi)$ which has a local minimum at $\phi=\phi_{\text{false}}$, a true minimum at $ \phi_{\text{true}}$, and a local maximum at $\phi_{\text{top}}$ as shown in FIG. \ref{fig:potential_schematic}.
We use the saddle point method and evaluate the tunneling rate by computing the action for the tunneling configurations \cite{PhysRevE.70.041106}. 
We investigate both HM and CDL tunneling processes.
In the case of a shallow potential barrier, the conventional HM tunneling rate is reproduced. 
Remarkably, the tunneling configuration in the MSRJD functional integral method has physically natural boundary conditions in contrast to the Euclidean method. We then clarify that the HM tunneling rate is the tunneling rate of a (roughly Hubble-sized) coarse-grained patch transitioning from $\phi_\text{false}$ to $\phi_\text{top}$. In the case of 
a steep potential barrier, we find an interesting CDL bubble-like configuration which is obtained by solving the solution of a scalar field equation in Euclidean anti-de Sitter space 
even though we do not work in the imaginary time. Our results show how the bubble nucleation process could be described in the stochastic approach.

This paper is organized as follows. 
In sec.~\ref{sec:Stochastic_approach}, we present the setup
and  review the stochastic approach.
In sec.~\ref{sec:pathint}, the MSRJD functional integral is constructed from the Langevin equations. 
We also outline the derivation of MSRJD functional integral starting from the Schwinger-Keldysh formalism. 
In sec.~\ref{sec:HM}, we apply the MSRJD functional integral to the case of the shallow potential barrier and study the HM tunneling process.  
In sec.~\ref{sec:CDL}, we consider the case of the steep potential barrier and find the CDL bubble-like configuration.  We derive the tunneling rate using the configuration and compare it with the one obtained in sec.~\ref{sec:HM}. We also compare our results with those in the Euclidean method. 
Sec.~\ref{sec:concl} is devoted to the conclusion. Some technical details are presented in appendices.

\section{Stochastic approach}
\label{sec:Stochastic_approach}
In this section, we review the stochastic approach \cite{Starobinsky:1986fx,Starobinsky:1994bd}.

\subsection{Setup}
We consider a real scalar field $\phi$ 
in (3+1) dimensional de Sitter background.
The background geometry is described by the metric
\begin{align}
\dm s^{2} 
= -\mathrm{d}t^{2}+a(t)^{2}\dm\bm{x}^{2} 
= a(\eta^{2})(-\dm\eta^{2}+\dm\bm{x}^{2}),\label{eq:Metric}
\end{align}
with the scale factor 
\begin{equation}
a(t) = e^{Ht} = -\frac{1}{H\eta}, 
\end{equation}
where $H$ is the Hubble constant.
The action is given by
\begin{equation}
S
=\int \mathrm{d}t\mathrm{d}^{3} x \, a^{3}
\bigg[\frac{1}{2}\dot{\phi}^{2} -\frac{1}{2} a^{-2} (\nabla\phi)^{2} - V(\phi)\bigg],
\end{equation}
where a dot denotes a derivative respect to $t$
and $V(\phi)$ is a potential function. From this action,  the canonical conjugate momentum $\Pi$ is defined as $\Pi:=a^3\dot\phi$
and the Hamilton's equations of motion are given by
\begin{equation}
\dot{\phi} = a^{-3}\Pi \ , \qquad
\dot{\Pi} =a \nabla^{2}\phi-a^{3}V'(\phi)
\,. \label{eq:HamiltonEq}
\end{equation}

\subsection{Langevin equation}
First, we divide the quantum fields $\phi$ and $\Pi$ into an infrared (IR) part and an ultraviolet (UV) part respectively as
\begin{equation}
\phi =\phi_{\text{IR}}+\phi_{\text{UV}} \ , \qquad
\Pi =\Pi_{\text{IR}}+\Pi_{\text{UV}}\ . \label{eq:IRUV}
\end{equation}
The UV part is defined as
\begin{align}
\phi_{\text{UV}}&:=\int \frac{\mathrm{d}^{3}k}{(2\pi)^{3}}\theta(k-k_{c}(t))
\bigg[
a_{\bm{k}}u_{\bm k}(t)e^{i\bm{k}\cdot\bm{x}} +
a_{\bm{k}}^{\dagger}u_{\bm k}^{*}(t)e^{-i\bm{k}\cdot\bm{x}}
\bigg], \notag\\
\Pi_{\text{UV}}&:=\int \frac{\mathrm{d}^{3}k}{(2\pi)^{3}}\theta(k-k_{c}(t))
a(t)^{3}\,\bigg[
a_{\bm{k}}\dot{u}_{\bm k}(t)e^{i\bm{k}\cdot\bm{x}} +
a_{\bm{k}}^{\dagger}\dot{u}_{\bm k}^{*}(t)e^{-i\bm{k}\cdot\bm{x}}
\bigg], \label{eq:UVpart}
\end{align}
where $u_{\bm k}$ is a mode function that will be specified later, and $a_{\bm{k}}$ and $a^{\dagger}_{\bm{k}}$ are annihilation and creation operators, respectively. Here, $\theta(k-k_{c}(t))$ is a step function and $k_{c}(t)$ is the cutoff scale defined as
\begin{equation}
k_{c}(t):=\varepsilon a(t)H,  \label{eq:cutoff}
\end{equation}
where $\varepsilon$ is a constant. The value of $\varepsilon$ will be discussed later.
Substituting Eqs.\eqref{eq:UVpart} into Hamilton's equations \eqref{eq:HamiltonEq}, we obtain the following equations;
\begin{equation}
\dot{\phi}_{\text{IR}}
=a^{-3}\Pi_{\text{IR}}+\xi^{\phi} \ , \qquad
\dot{\Pi}_{\text{IR}}
=a\nabla^{2}\phi_{\text{IR}}-a^{3}V'(\phi_{\text{IR}})+\xi^{\Pi} \ , \label{eq:LangevinFull}
\end{equation}
where  $\xi^{\phi}$ and $\xi^{\Pi}$ are defined as
\begin{align}
\xi^{\phi}&:=\dot{k}_{c}(t)\int \frac{\mathrm{d}^{3}k}{(2\pi)^{3}}\delta(k-k_{c}(t))
\bigg[
a_{\bm{k}}u_{\bm k}(t)e^{i\bm{k}\cdot\bm{x}} +
a_{\bm{k}}^{\dagger}u_{\bm k}^{*}(t)e^{-i\bm{k}\cdot\bm{x}}
\bigg], \notag\\
\xi^{\Pi}&:=\dot{k}_{c}(t)\int \frac{\mathrm{d}^{3}k}{(2\pi)^{3}}\delta(k-k_{c}(t))
\bigg[
a_{\bm{k}}a(t)^{3}\dot{u}_{\bm k}(t)e^{i\bm{k}\cdot\bm{x}} +
a_{\bm{k}}^{\dagger}a(t)^{3}\dot{u}_{\bm k}^{*}(t)e^{-i\bm{k}\cdot\bm{x}}
\bigg]. \label{eq:noise}
\end{align}
To derive Eqs.\eqref{eq:LangevinFull} and \eqref{eq:noise}, 
we assumed that the time evolution of UV modes can be well approximated by the free theory that is defined around the false vacuum. 
The mode function $u_{\bm k}(t)$ introduced in \eqref{eq:noise} then satisfies the following equation
\begin{align}
\ddot{u}_{\bm k}+3H\dot{u}_{\bm k} + a^{-2}k^{2}u_{\bm k} 
+ V''(\phi_\text{false})u_{\bm k} = 0.
\label{eq:ModeEq}
\end{align}
From this equation, we see the mode function only depends on $k:=|\bm{k}|$. Hence, we simply denote the mode function as $u_{k}$.
Note that this mode function is used for UV modes $k>k_{c}(t)$.

Let us consider statistical properties of $\xi^{\phi}$ and $\xi^{\Pi}$.
The annihilation and creation operators $a_{\bm{k}}$ and $a^{\dagger}_{\bm{k}}$ satisfy following commutation relations:
\begin{align}
[a_{\bm{k}}, a^{\dagger}_{\bm{k}'}]=(2\pi)^{3}\delta^{(3)}(\bm{k}-\bm{k}') \ , 
\qquad
[a_{\bm{k}}, a_{\bm{k}'}]=[a^{\dagger}_{\bm{k}}, a^{\dagger}_{\bm{k}'}]=0.\label{eq:commutation}
\end{align}
Defining the vacuum state $\ket{0}$ as
\begin{align}
a_{\bm{k}}\ket{0}=0, \quad \forall \bm{k}, \label{eq:vacuum}
\end{align}
we can calculate the one point and two point correlation functions of $\xi^{\phi}$ and $\xi^{\Pi}$ as
\begin{align}
&\braket{0|\xi^{\alpha}(t,\bm{x})|0}
=0, \quad (\alpha,\beta= \phi, \Pi),  \notag\notag\\
&\braket{0|\xi^{\alpha}(t,\bm{x})\xi^{\beta}(t',\bm{x}')|0}
=\frac{1}{2\pi^{2}}\dot{k}_{c}(t)k_{c}(t)^{2}\, \frac{\sin(k_{c}(t)r)}{k_{c}(t)r}g^{\alpha\beta}(t)\delta(t-t'), \label{eq:correlation}
\end{align}
where $r:=|\bm{x}-\bm{x}'|$ and
\begin{align}
\begin{cases}
g^{\phi\phi}(t)&:=|u_{k_{c}}(t)|^{2}\\
g^{\Pi\Pi}(t)&:=a(t)^6|\dot{u}_{k_{c}}(t)|^{2}\\
g^{\phi\Pi}(t)&=(g^{\Pi\phi})^{*}:=a(t)^{3}u_{k_{c}}(t)\dot{u}_{k_{c}}^{*}(t).
\end{cases}. \label{eq:Function_g}
\end{align}
Because of the Wick theorem, all higher correlation functions can be decomposed into two point functions. 
In the stochastic approach, we replace the operator $\xi^{\alpha}$ by the real random field which has the following statistical properties
\begin{align}
&\braket{\xi^{\alpha}(t,\bm{x})}_{\xi}
=0, \quad (\alpha,\beta= \phi, \Pi),  \notag\notag\\
&\braket{\xi^{\alpha}(t,\bm{x})\xi^{\beta}(t',\bm{x}')}_{\xi}
=
\frac{1}{2\pi^{2}}\dot{k}_{c}(t)k_{c}(t)^{2}\, \frac{\sin(k_{c}(t)r)}{k_{c}(t)r} \text{Re}[g^{\alpha\beta}(t)]\delta(t-t'),
\label{symmetric_noise}
\end{align}
where $\braket{\cdots}_{\xi}$ represents an expectation value with the distribution function of $\xi^{\alpha}$.
With this prescription, Eqs.\eqref{eq:LangevinFull} can be interpreted as the Langevin equations with the noise $\xi^{\alpha}$ stemming from quantum fluctuations.
Since $\xi^{\alpha}$ are now regarded as classical variables, we ignored the imaginary parts of $g^{\phi\Pi}$ and $g^{\Pi\phi}$ which come from non-commutativity.
Accordingly, IR variables $\phi_\text{IR}$ and $\Pi_\text{IR}$ are also regarded as classical stochastic variables in this prescription. The validity of this classical approximation will be discussed in the next section.

\subsection{Bunch-Davies vacuum}
Now we define the vacuum state $\ket{0}$ by specifying the mode function $u_k$. Eq.\eqref{eq:ModeEq} is the equation for the mode $k>k_{c}$. Let us assume that we can ignore the term $V''$ in solving \eqref{eq:ModeEq}. This can be justified when $|V''|$ is sufficiently small to satisfy $\max\left[(k_{c}/a)^2, H^2\right]\gg |V''|$. 
Under this approximation, \eqref{eq:ModeEq} can be solved as 
\begin{align}
u_{k}=(-k\eta)^{\frac{3}{2}}\bigg[C_{1}H^{(1)}_{\frac{3}{2}}(-k\eta)+C_{2}H^{(2)}_{\frac{3}{2}}(-k\eta)\bigg],
\end{align}
where $C_{1}, C_{2}$ are constants of integration and $H^{(1)}_{\nu}(z), H^{(2)}_{\nu}(z)$ are Hankel functions.
We choose the Bunch-Davies vacuum, i.e., the vacuum coincides with Minkowski vacuum at past infinity ($\eta \to -\infty$) so that
\begin{align}
u_{k} \to \frac{-H\eta}{\sqrt{2k}}e^{-ik\eta}, \quad (-k\eta \to \infty). \label{eq:MinkowskiLimit}
\end{align}
This determines $C_{1}, C_{2}$ as
\begin{align}
C_{1}=-\sqrt{\frac{\pi}{4k}}\frac{H}{k} \ ,  \qquad
C_{2}=0 \ . \label{eq:coefficients}
\end{align}
Thus, we obtain the mode function
\begin{align}
u_{k}
= \frac{H\eta}{\sqrt{2k}}\bigg(\frac{1}{-ik\eta}-1\bigg)e^{-ik\eta}. \label{eq:ModeFunctionExplit}
\end{align}
To check \eqref{eq:MinkowskiLimit}, we used the following formulae
\begin{align}
H^{(1)}_{\frac{3}{2}}(z) 
= \sqrt{\frac{2}{\pi z}}\bigg(\frac{1}{iz}-1\bigg)e^{iz}\ , \qquad
H^{(2)}_{\frac{3}{2}}(z) 
= \sqrt{\frac{2}{\pi z}}\bigg(\frac{1}{-iz}-1\bigg)e^{-iz}.
\end{align}
Substituting \eqref{eq:ModeFunctionExplit} into \eqref{eq:Function_g} and using \eqref{eq:cutoff}, we obtain
\begin{align}
&\begin{cases}
g^{\phi\phi}&=\frac{H^{2}\eta^{2}}{2k_{c}}\bigg(\frac{1}{k_{c}^{2}\eta^{2}}+1\bigg)
=\frac{1}{2Ha^{3}}\bigg(\varepsilon^{-3} + \varepsilon^{-1}\bigg)\\
g^{\Pi\Pi}&=\frac{k_{c}}{2H^{2}\eta^{2}}
=\frac{Ha^{3}}{2}\varepsilon\\
g^{\phi\Pi}&=g^{\Pi\phi*}=\frac{1}{2}\bigg(\frac{1}{k_{c}\eta}+i\bigg)
=-\frac{1}{2}\varepsilon^{-1} + \frac{i}{2}
\end{cases}. \label{eq:Function_g_Bunch}
\end{align}
From \eqref{eq:LangevinFull}, we have to scale $g^{\alpha\beta}$ like as $g^{\phi\phi} \to g^{\phi\phi}$, $g^{\Pi\Pi} \to g^{\Pi\Pi}/(H^2a^6)$ and $g^{\phi\Pi} \to g^{\phi\Pi}/(Ha^3)$ in order to compare these quantities.
Then,  
$g^{\phi\phi}$ and $g^{\Pi\Pi}$ become dominant for the case $\varepsilon \ll 1$ and $\varepsilon \gg 1$, respectively.

\subsection{Reduced Langevin equation}
Let us focus on physics at the super-horizon scales by taking $\varepsilon\ll1$ and assume a shallow potential so that we can use the massless approximation to evaluate the noise correlations. 
In this setup, we can use \eqref{eq:Function_g_Bunch} from which we find that 
$g^{\phi\phi}$ becomes dominant. 
We then ignore the noise for momentum $\xi^\Pi$.
We also ignore the gradient term in \eqref{eq:LangevinFull} because we focus on the physics at super-horizon scales. Since the potential is assumed to be shallow, the time variation of $\phi_{\text{IR}}(t)$ will be then very small. Therefore, we may integrate the second equation of \eqref{eq:LangevinFull} as 
\begin{align}
\Pi_{\text{IR}} \simeq -\frac{a^{3}}{3H}V'(\phi_{\text{IR}}) \label{eq:slow-roll}.
\end{align}
Substituting this relation into the first equation of \eqref{eq:LangevinFull}, we obtain
\begin{align}
\dot{\phi}_{\text{IR}} = -\frac{V'(\phi_{\text{IR}})}{3H} + \xi^{\phi},\label{eq:LangevinSY}
\end{align}
where the two point correlation function of $\xi^{\phi}$ is given at the leading order by
\begin{align}
    \braket{\xi^{\phi}(t,\bm{x}) \xi^{\phi}(t',\bm{x}')}_{\xi} 
    =
    \frac{H^{3}}{4\pi^{2}} j_0(k_cr)\delta(t-t')
    \,,
    \quad
    j_0(k_cr)
    \coloneqq
    \frac{\sin(k_{c}r)}{k_{c}r}
    \,.\label{eq:CorrelationSY}
\end{align}
Eq.\eqref{eq:LangevinSY} is the Langevin equation with the noise correlation \eqref{eq:CorrelationSY} derived in \cite{Starobinsky:1994bd}.

\section{Path integral formalism for the stochastic approach
}\label{sec:pathint}

In this section, we introduce path integral approach for the Langevin equation following \cite{altland_simons_2010}.
From now on, we replace the IR field and its conjugate momentum as $(\phi_{\text{IR}},\Pi_{\text{IR}}) \to (\phi_{c},\Pi_{c})$.
As we see in section~\ref{sec:SK_transition}, this notation corresponds to the Keldysh basis. In section~\ref{sec:SK_transition}, we provide a first-principle derivation of the stochastic approach based on the Schwinger-Keldysh formalism following~\cite{Tokuda:2017fdh,Tokuda:2018eqs}.\footnote{For earlier discussions of stochastic approach based on the Schwinger-Keldysh formalism, see e.g., \cite{Morikawa:1989xz,Tolley:2008qv}.} We briefly show the sketch of the derivation in the main text, and details of derivations are given in appendix~\ref{sec:in-in}. 

\subsection{MSRJD functional integral representation}
\subsubsection{0+1 dimensional theories}
Eq.\eqref{eq:LangevinSY} can be regarded as the equation determining the position of a particle  $\phi(t)$ with the white noise $\xi^{\phi}$ defined by Eq.\eqref{eq:CorrelationSY}. 
For $k_c r \ll 1$, we have $\sin(k_{c}r)/(k_{c}r) \simeq 1$ in a good approximation. Let us discretize the field and noise as $\phi_c(t) \to \phi_{i}, \, \xi^{\phi}(t) \to \xi^{\phi}_{i}, \, (i \in \mathbf{Z})$.
Then,  Eq.\eqref{eq:LangevinSY} reads
\begin{align}
X_{i}:= \phi_{i}-\phi_{i-1} + \Delta t \bigg(\frac{V'(\phi_{i})}{3H} - \xi^{\phi}_{i}\bigg) = 0, \label{eq:discritization}
\end{align}
where $\Delta t$ is the temporal discretization interval. 
Note that the discretization \eqref{eq:discritization} is called Ito discretization.\footnote{Generally, we can choose other discretization like as Stratonovich discretization. 
This ambiguity does not affect the result since the amplitudes of the noise do not depend on fields within the range of our approximation.}
Denoting a solution of \eqref{eq:LangevinSY} by $\phi_c[\xi]$, the expectation value of an observable $\braket{\mathcal{O}}_{\xi}$ can be formally expressed as
\begin{align}
\braket{\mathcal{O}[\phi_c(t)]}_{\xi}
= \int \mathcal{D}\phi_c \, \mathcal{O}[\phi_c] \braket{\delta(\phi_c-\phi_c[\xi])}_{\xi} 
= \int \mathcal{D}\phi_c \, \mathcal{O}[\phi_c] \Braket{\bigg| \frac{\delta X}{\delta \phi_c}\bigg| \, \delta(X)}_{\xi}. \label{eq:ExpectationFormal}
\end{align}
where $\mathcal{D}\phi_c := \prod_{i} d\phi_{i}$ is the functional measure, $\delta(\phi_c-\phi_c[\xi]) := \prod_{i} \delta(\phi_{i}-\phi_{i}[\xi])$, and $|\delta X / \delta \phi_c|$ is the determinant of $\{\delta X_{i} / \delta \phi_{j}\}$.
 When we choose the above discretization \eqref{eq:discritization}, $\delta X / \delta \phi_c$ becomes triangular matrix with a unit in the diagonal components, i.e., the functional determinant is unity.
Thus, substituting \eqref{eq:discritization} into \eqref{eq:ExpectationFormal}, we obtain
\begin{align}
\braket{\mathcal{O}[\phi_c(t)]}_{\xi}
= \int \mathcal{D}\phi_c \, \mathcal{O}[\phi_c]
\Braket{\prod_{i} \delta\bigg(\phi_{i}-\phi_{i-1} + \Delta t \bigg(\frac{V'(\phi_{i})}{3H} - \xi^{\phi}_{i}\bigg)\bigg)}_{\xi}.
\end{align}
Representing the delta functions in terms of a Fourier integral and taking the continuum limit, we obtain the following path integral representation
\begin{align}
\braket{\mathcal{O}[\phi_c(t)]}_{\xi}
= \int \mathcal{D}(\phi_c, \tilde{\phi}) \, \mathcal{O}[\phi_c]
\Braket{
\exp\bigg[i\int \mathrm{d}t \tilde{\phi} \bigg(\dot{\phi}_c + \frac{V'(\phi_c)}{3H} -\xi^{\phi} \bigg)\bigg] }_{\xi}.
\end{align}
After averaging over the noise assuming Gaussian statistics, we obtain the Martin–Siggia–Rose –Janssen–de Dominicis (MSRJD) functional integral
\begin{align}
\braket{\mathcal{O}[\phi_c(t)]}_{\xi}
= \int \mathcal{D}(\phi_c, \tilde{\phi}) \, \mathcal{O}[\phi_c]
\exp\bigg[\int \mathrm{d}t \bigg(
i\tilde{\phi} \bigg(\dot{\phi}_c + \frac{V'(\phi_c)}{3H}\bigg) 
- \frac{H^{3}}{8\pi^{2}}\tilde{\phi}^{2} \bigg)\bigg].
\end{align}
Putting $\mathcal{O}[\phi_c] = \delta(\phi_c - \phi_c(t))$ and taking the boundary conditions $\phi_c(t')=\phi_c'$ and $\phi_c(t)=\phi_c$, we obtain the transition probability $p(\phi_c, t | \phi_c', t')$ 
\begin{align}
p(\phi_c, t | \phi_c', t')
=\int_{\phi_c(t')=\phi_c'}^{\phi_c(t)=\phi_c} \mathcal{D}(\phi_c, \tilde{\phi}) \,
\exp\bigg[\int \mathrm{d}t \bigg(
i\tilde{\phi} \bigg(\dot{\phi}_c + \frac{V'(\phi_c)}{3H}\bigg) 
- \frac{H^{3}}{8\pi^{2}}\tilde{\phi}^{2} \bigg)\bigg]
\,.\label{eq:onedim_original}
\end{align}
Finally, introducing new variables 
$\Pi_\Delta\coloneqq i\tilde\phi$,
we obtain the "phase space" path integral 
\begin{eqnarray}
p(\phi_c, t | \phi_c', t')
=\int_{\phi_c(t')=\phi_c'}^{\phi_c(t)=\phi_c} \mathcal{D}(\phi_c, \Pi_{\Delta}) \, 
\exp\bigg[\int \mathrm{d}t (\Pi_{\Delta}\dot{\phi}_c - H(\phi_c, \Pi_{\Delta}) )\bigg]
\ , 
\label{eq:ConditionalSY}
\end{eqnarray}
where we defined the Hamiltonian as
\begin{eqnarray}
H(\phi_c, \Pi_{\Delta}):= -\frac{V'(\phi_c)}{3H}\Pi_{\Delta} - \frac{H^{3}}{8\pi^{2}}\Pi_{\Delta}^{2} \ .
\label{eq:def_Hamiltonian_HM}
\end{eqnarray}
We emphasize that this is just a change of integration variable, not the Wick rotation. 
A set of saddle point equations is defined as the Hamilton's equations for the Hamiltonian defined above.

\subsubsection{3+1 dimensional theories}
Let us consider the case of the full Langevin Eqs. \eqref{eq:LangevinFull}, \eqref{eq:correlation} and \eqref{eq:Function_g_Bunch}.
When the space is discretized, \eqref{eq:LangevinFull} can be seen as the dynamics of many body particles. Thus, the procedure in the previous subsection can be utilized. Taking the continuum limit, we obtain the following expression which is analogous to \eqref{eq:onedim_original},
\begin{align}
p(\phi_c(\bm{x}),t |\phi_c'(\bm{x}), t')
&=\int_{\phi_c(t',\bm{x})=\phi_c'(\bm{x})}^{\phi_c(t,\bm{x})=\phi_c(\bm{x})} \mathcal{D}(\phi_c, \Pi_c, \phi_{\Delta}, \Pi_{\Delta}) \notag\\
&\times \exp\bigg[ i\int \mathrm{d}^{4}x 
\bigg(
\Pi_{\Delta}\dot{\phi}_c - \phi_{\Delta}\dot{\Pi}_c - H(\phi_c, \Pi_c, \phi_{\Delta}, \Pi_{\Delta})
\bigg)\bigg] \ . \quad \notag\\
\label{eq:ConditionalFull}
\end{align}
We defined the Hamiltonian $H(\phi_c, \Pi_c, \phi_{\Delta}, \Pi_{\Delta})$ as
\begin{align}
H(\phi_c,\Pi_c,\phi_{\Delta},\Pi_{\Delta}) 
=
 &\frac{\Pi_c \Pi_{\Delta}}{a^3} -(a\nabla^{2}\phi_c - a^{3}V'(\phi_c))\phi_{\Delta}
 \no\\
 &
-\frac{i}{2}\sum_{\alpha,\beta}\int \mathrm{d}^{4}x'  X_{\alpha}(x)G^{\alpha\beta}(x,x')X_{\beta}(x')\,,\label{eq:def_Hamiltonian}
\end{align}
where the Greek indices $\alpha,\beta=(\phi,\Pi)$ label the $\Delta$ fields as $(X_\phi,X_\Pi)=(\Pi_\Delta, -\phi_\Delta)$ and $G^{\alpha\beta}$ denotes the correlations of noises: 
\begin{equation}
    G^{\alpha\beta}(x,x')
    \coloneqq \langle\xi^\alpha(x)\xi^\beta(x')\rangle_\xi
    \,,\label{eq:defG}
\end{equation}
where the RHS is given in \eqref{symmetric_noise}.
We use the formula \eqref{eq:ConditionalFull} in sections~\ref{subsec:HM_full} and \ref{sec:CDL}. 
Note that we have not changed the integration variables as $(\phi_\Delta,\Pi_\Delta)\to(-i\phi_\Delta,-i\Pi_\Delta)$.

In sections~\ref{sec:HM} and \ref{sec:CDL}, we take the initial state and the final state to be the false vacuum and the field configurations after tunneling (true vacuum or bubble), respectively.

\subsection{Transition probability from the Schwinger-Keldysh formalism}\label{sec:SK_transition}
In the above prescription, we treat IR fields $(\phi_\text{IR},\Pi_\text{IR})$ as classical stochastic variables.  
To quantify the validity of this approximation, 
we derive the stochastic approach from the first principles based on the Schwinger-Keldysh formalism. Note that we recover the label ``IR'' for IR fields in this subsection following the notation adopted in appendix~\ref{sec:in-in}.

Suppose that the system is in the false vacuum
$\ket{\Psi_\text{false}}$ at the past infinity $t=-\infty$ and it evolves into a certain quantum state $|\{\phi^\text{IR}(T,\bm{x})\}_{\bm x\in\mathcal{D}}\rangle$ at the final time $t=T$, where $|\{\phi^\text{IR}(T,\bm{x})\}_{\bm x\in\mathcal{D}}\rangle$ is an eigenstate of an IR field $\hat\phi^\text{IR}(\bm x)$ at spatial points $\bm x$ in the domain $\mathcal{D}$ with an eigenvalue $\phi^\text{IR}(T,\bm x)$. Here, $\hat\phi^\text{IR}(\bm{x})$ is an IR field in the Schr\"{o}dinger picture.\footnote{Precisely, $\hat\phi^\text{IR}(\bm x)$ is defined in terms of the the Schr\"{o}dinger picture field $\hat\phi({\bm k})$ in momentum space as 
$
    \hat\phi^\text{IR}(\bm x)
    \coloneqq
    \int\frac{\mathrm{d}^3k}{(2\pi)^3}\,\hat\phi({\bm{k}})\,e^{i\bm{k}.\bm{x}}\theta(k_c(t)-k)
$. We suppressed the trivial time dependence stemming from the step function.
} 
The transition probability $p$ for this process is given by
\begin{equation}
    p(\{\phi^\text{IR}(T,\bm{x})\}_{\bm x\in\mathcal{D}})
    =
    \left|\langle\{\phi^\text{IR}(T,\bm{x})\}_{\bm x\in\mathcal{D}}|\hat U(t,-\infty)|\Psi_\text{false}\rangle
    \right|^2
    \,,
\end{equation}
where $\hat U(t,t')$ describes the unitary time evolution from $t'$ to $t$. The RHS is the Fourier component of the generating functional $Z[J^\text{IR}]$ for IR fields, and we have
\begin{align}
    &p(\{\phi^\text{IR}(T,\bm{x})\}_{\bm x\in\mathcal{D}})=
    \left.
    \prod_{\bm x\in\mathcal{D}}
    \int\mathrm{d}J^\text{IR}(T,\bm x)
     Z[J^\text{IR}(T)]
     e^{-iJ^\text{IR}(T,\bm x)\phi^\text{IR}(T,\bm x)}
     \right|_{\{J^\text{IR}(T,\bm y)=0\}_{\bm y\notin \mathcal{D}}}
     \,,\label{tunnel_IRgene}
\end{align}
where the generating functional $Z[J^\text{IR}(T)]$ is defined by 
\begin{align}
&    Z[J^\text{IR}(T)]
    \coloneqq
    \text{Tr}
    \left[
        \hat U(T,-\infty)\ket{\Psi_\text{false}}\bra{\Psi_\text{false}}\hat U^\dagger(T,-\infty)
         \prod_{\bm x}e^{iJ^\text{IR}(T,\bm x)\phi^\text{IR}(T,\bm x)}
    \right]
    \,.
\label{IRgene_def}
\end{align}

We can evaluate $Z[J^\text{IR}(T)]$ non-perturbatively in the IR sector based on the method developed in \cite{Tokuda:2017fdh,Tokuda:2018eqs}. Our strategy is that we integrate out UV sector $k>k_c(t)$ perturbatively in nonlinear couplings to obtain $Z[J^\text{IR}(T)]$. This is analogous to tracing out environmental degrees of freedom to evaluate the reduced density matrix for the system under considerations. 
Each step we take can be summarized as follows:
\begin{enumerate}
    \item First, we derive a path integral representation of $Z[J^\text{IR}(T)]$. We then split the integration variables into UV fields and IR fields such that the integration contour of UV variables of $\phi$ is closed: see discussions around Eq.\eqref{repl_phi} for more details.\footnote{There is a subtlety that modes satisfying $k\geq k_c(T)$ are initially regarded as the ``UV'' degrees of freedom (DoF) while they become ``IR'' DoFs due to the accelerating expansion of the spacetime. However, by adopting this splitting procedure, we can use the Schwinger-Keldysh (or {\it closed-time-path}) formalism to evaluate the integration over UV variables first as usual.}
    \item We perform the integration over UV variables and evaluate an IR effective action, the so-called Feynman-Vernon influence functional~\cite{Feynman:1963fq} perturbatively.
    \item We substitute the obtained expression of $Z[J^\text{IR}(T)]$ into \eqref{tunnel_IRgene}, giving rise to the path integral expression for the transition probability $p$.
\end{enumerate}
Technical details are shown in appendix~\ref{sec:in-in}. At the step 2, 
we assumed that the quantum state for non-zero modes is given by the Bunch-Davies vacuum state for a free field defined around the false vacuum, and that the exact zero mode provides a non-fluctuating classical background field configuration $\phi=\phi_\text{false}$.

In this way, we can obtain the path integral expression for the probability $p$ as
\begin{align}
    p(\{\phi^\text{IR}(T,\bm{x})\}_{\bm x\in\mathcal{D}})=
    &\int\mathcal{D}(\phi^\text{IR}_c,\phi^\text{IR}_\Delta,\Pi^\text{IR}_c,\Pi^\text{IR}_\Delta)
    \,
    \,e^{i\mathcal{S}}\, \delta\left[\mathcal{C}_\text{fin}\right]\delta\left[\mathcal{C}_\text{ini}\right]
   \,.\label{tunnel_IRgene_main}
\end{align}
Here, $\delta\left[\mathcal{C}_\text{fin}\right]$ and $\delta\left[\mathcal{C}_\text{ini}\right]$ fix the boundary conditions of path integral at the final time and the initial time, respectively:
\begin{subequations}
\label{eq:pathint_bc}
\begin{align}
    &\delta\left[\mathcal{C}_\text{fin}\right]
    \coloneqq
    \prod_{\bm x\in\mathcal{D}}\delta(\phi^\text{IR}_c(T,\bm x)-\phi^\text{IR}(T,\bm x))
    \prod_{\bm y}
    \delta(\phi^\text{IR}_\Delta(T,\bm y))
    \,,\label{pathint_bc_fin}\\
    &\delta\left[\mathcal{C}_\text{ini}\right]
    \coloneqq
    \prod_{\bm x}\delta
    \left(
        \phi^\text{IR}_c(-\infty,\bm x)
        -\phi_\text{false}
    \right)
    \delta\left(\Pi^\text{IR}_c(-\infty,\bm x)\right)\delta[\mathcal{C_\text{detail}}]
    \,,\label{pathint_bc_ini}
\end{align}    
\end{subequations}
where $\delta[\mathcal{C}_\text{detail}]$ is not relevant for the main points here, and we define it in appendix~\ref{sec:in-in}.
A term $\mathcal{S}$  
may be understood as an IR effective action,
\begin{subequations}
\label{eq:effective_action}
\begin{align}
    &\mathcal{S}
    \coloneqq
    \int\mathrm{d}^4x\,
    \left[
    \Pi^\text{IR}_\Delta\dot\phi^\text{IR}_c
    -\phi^\text{IR}_\Delta \dot\Pi^\text{IR}_c
    - H(\phi^\text{IR}_c,\Pi^\text{IR}_c,\phi^\text{IR}_\Delta,\Pi^\text{IR}_\Delta)
    -\delta H
    \right]
    \,,
    \\
    &\delta H
    \coloneqq
    a^3(t)\left[
        V(\phi^\text{IR}_c+(\phi^\text{IR}_\Delta/2))
        -
        V(\phi^\text{IR}_c-(\phi^\text{IR}_\Delta/2))
        -
        V'(\phi^\text{IR}_c)\phi^\text{IR}_\Delta
    \right]
    +\delta H_\text{higher}
    \,,
\end{align}    
\end{subequations}
where $H$ is given in \eqref{eq:def_Hamiltonian}. $\delta H_\text{higher}$ denotes the higher-order terms in the coupling constant.
Note that eq~\eqref{tunnel_IRgene_main} allows us to understand the transition probability $p$ in terms of the stochastic dynamics: this point is discussed in appendix~\ref{sec:interpretation}.

\paragraph{Validity of Eq.\eqref{eq:ConditionalFull}.} Now we find that $\mathcal{S}$ coincides with the term in the exponent of \eqref{eq:ConditionalFull} up to the term $\delta H$.
Hence, \eqref{tunnel_IRgene_main} shows that the result \eqref{eq:ConditionalFull} in the previous section can be justified from the first-principles when the term $\delta H$ is negligible.
Suppose that the potential $V(\phi)$ is sufficiently flat in the regime of our interest such that perturbation theory works when integrating out UV modes. In this case, we can systematically calculate $\delta H_\text{higher}$ perturbatively and we have $\delta H_\text{higher}=0$ at the leading order. We set $\delta H_\text{higher}=0$ in the present analysis. We should keep in mind however that the non-perturbative physics at UV scales $k> k_c(t)=\varepsilon a(t)H$ is lost by setting $\delta H_\text{higher}=0$. From this viewpoint, we should choose $\varepsilon$ as large as possible.

Even after setting $\delta H_\text{higher}=0$, we have nonzero $\delta H$. In the classical approximation, we may regard the $\Delta$-variables $(\phi^\text{IR}_\Delta,\Pi^\text{IR}_\Delta)$ as tiny quantities and keep only the terms which are linear in the $\Delta$-variables in $\mathcal S$, leading to $\delta H\simeq 0$. In the case $\varepsilon\ll1$, this approximation would work thanks to the squeezing of quantum fluctuations at super-horizon scales. However, it is more subtle if we can set $\delta H=0$ in the case $\varepsilon \gtrsim 1$. We will revisit this issue in section~\ref{sec:CDL}.

\section{Hawking-Moss tunneling}\label{sec:HM}
In this section, we discuss the Hawking-Moss (HM) tunneling from the perspective of MSRJD functional integral. 
We calculate the tunneling rate defined by \eqref{eq:ConditionalSY} and \eqref{eq:ConditionalFull} in sections~\ref{subsec:HM_onedim} and \ref{subsec:HM_full}, respectively.   
\begin{figure}[btp]
\centering
\includegraphics[width=100mm]{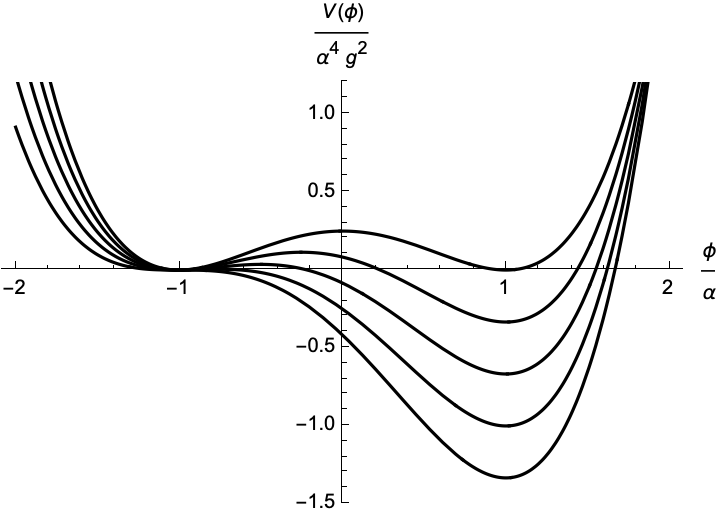}
\caption{The potential of \eqref{eq:potential} for various $\beta$. From top to bottom, the curves correspond to $\beta=0, 0.25, 0.5, 0.75, 1$.}
\label{fig:potential_eg}
\end{figure}
Hereafter, for the numerical calculations, we consider the  potential \cite{Weinberg:2012pjx} in FIG. \ref{fig:potential_eg};
\begin{align}
V(\phi) = \frac{g^{2}}{4}\bigg[ (\phi^{2}-\alpha^{2})^{2} + \frac{4\beta}{3} (\alpha\phi^{3} - 3\alpha^{3}\phi -2\alpha^{4})\bigg],
\label{eq:potential}
\end{align}
for which the derivative with respect to $\phi$ is given by a simple form
\begin{align}
    V'(\phi)
    =
    g^2(\phi-\alpha)(\phi+\alpha)(\phi+\alpha\beta)
    \,.
\end{align}
For this potential, we have $\phi_{\text{false}} = -\alpha$, $\phi_{\text{true}} = \alpha$ and $\phi_{\text{top}} = -\beta\alpha$.
The dimensionless parameter $\beta$ must be $0 < \beta <1$, otherwise $\phi = -\alpha$ is not a false vacuum. The potential at each point takes the following values:
\begin{align}
    &V(-\alpha) = 0\,, \quad 
    V(\alpha)= -\frac{4\beta g^2\alpha^4}{3}\,,\quad
    V(-\beta\alpha) = \frac{g^2\alpha^4}{12}(1-\beta)^3(3+\beta)\,.
\end{align}
The height of potential barrier $\Delta V$ between the false vacuum and the true vacuum is 
\begin{align}
\Delta V \coloneqq V(-\beta\alpha)-V(-\alpha) = \frac{g^2\alpha^4}{12}(1-\beta)^3(3+\beta)\,.
\end{align}
The mass of fluctuations around a given point $\phi$ is 
\begin{align}
    V''(\phi)
    =
    \alpha ^2 g^2 \left[3 (\phi/\alpha)^2+2 \beta  (\phi/\alpha)-1\right]\,.
\end{align}

In our analysis, we work in the quantum field theory on the fixed de Sitter background. This assumption will be valid when the condition $\max\bigl[|V(-\alpha)|,|V(-\beta\alpha)|,|V(\alpha)|\bigr] \ll 3M_\text{pl}^2H^2$ is satisfied. This is always satisfied when the following condition is imposed,
\begin{equation}
    H 
    \ll 
    \frac{3gM_\text{pl}}{2}\left(\frac{H}{g\alpha}\right)^2
    \approx
    2.9\times 10^{12}\,\text{GeV}
    \left(\frac{g}{1.12\times 10^{-4}}\right)
    \left(\frac{\sqrt{140}}{g\alpha/H}\right)^2
    \,,\label{eq:backreaction}
\end{equation}
where we use $0<\beta<1$ and choose parameters $(g,g\alpha/H)=(1.12\times 10^{-4},\sqrt{140})$ as a benchmark point for later convenience. This shows that we can ignore the backreaction consistently with the weak coupling $g\ll 1$ provided that $H$ is well below the Planck scale $M_\text{pl}\approx 2.44\times 10^{18}$ GeV.
%
Note that under this condition, our discussion below is also applicable in the case of a quasi de Sitter background. However, it is non-trivial to generalize our analysis to the case when a non-negligible backreaction on $H$ exists. This is because it is necessary to extend the first-principles derivation of the stochastic approach to account for the backreaction. This is in itself an interesting issue which we leave for future work.
%

\subsection{The case for the reduced Langevin equation}\label{subsec:HM_onedim}
\begin{figure}[btp]
\centering
\includegraphics[width=100mm]{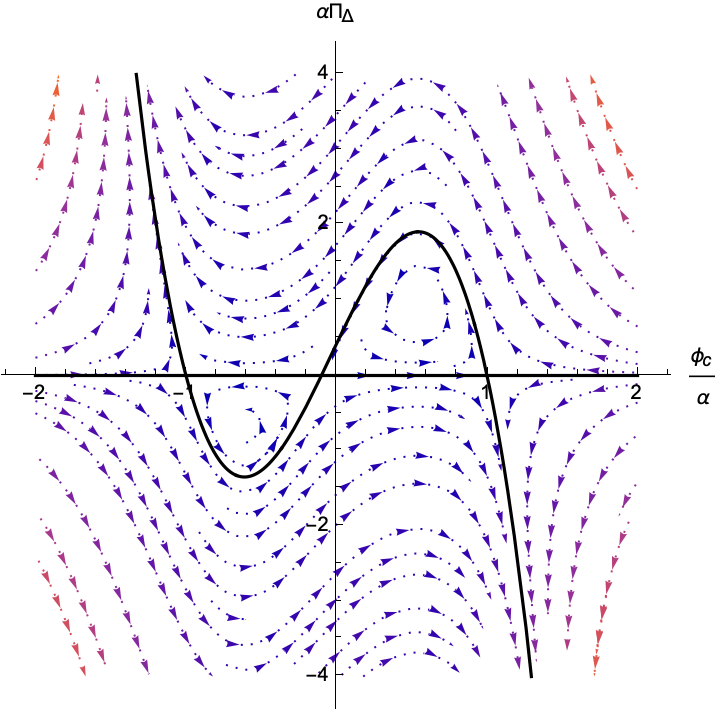}
\caption{The Hamilton flow for \eqref{eq:HamiltonEqSY} where we take $\alpha=H$, $g=0.4$ and $\beta=0.1$. The solid lines correspond to $H(\phi_c, \Pi_{\Delta})=0$. From left to right, the intersections of the solid lines correspond to the false vacuum, the top of potential hill and the true vacuum.}
\label{fig:HamiltonFlow}
\end{figure}

\begin{figure}[btp]
\centering
\includegraphics[width=100mm]{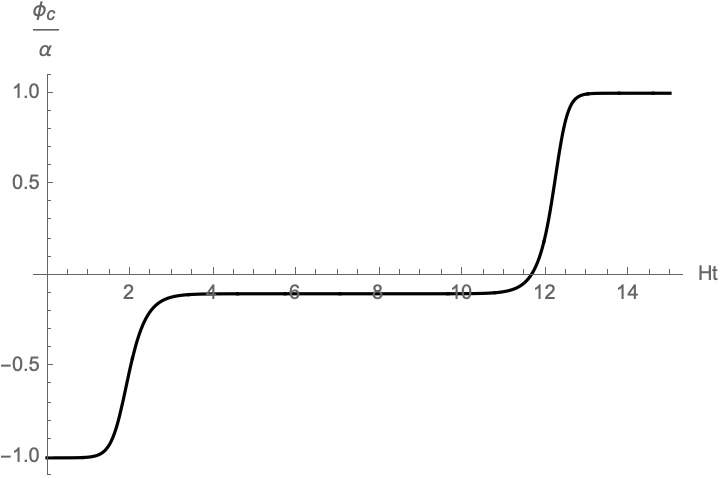}
\caption{The tunneling configuration for the homogeneous fields is plotted for $\beta=0.1$ and $\alpha g/H=\sqrt{10}$. We choose the initial condition as $\phi_c(10^{-15})/\alpha=-1+10^{-5}$ and solve the second equation of \eqref{eq:HamiltonEqSYZero}. When $\phi_c/\alpha$ reached the top of potential $\phi_c/\alpha = -\beta$, we switch to the first equation of \eqref{eq:HamiltonEqSYZero} with the initial condition $\phi_c/\alpha=-\beta+10^{-5}$.}
\label{fig:HomoBounce}
\end{figure}

Let us start from the simpler formula \eqref{eq:ConditionalSY}. 
The Hamilton's equations for the Hamiltonian $H(\phi_c,\Pi_\Delta)$ defined in \eqref{eq:def_Hamiltonian_HM} are given as
\begin{align}
\dot{\phi}_c &= -\frac{V'(\phi_c)}{3H} -\frac{H^{3}}{4\pi^{2}}\Pi_{\Delta} \notag\\
\dot{\Pi}_{\Delta} &= \frac{V''(\phi_c)}{3H}\Pi_{\Delta}. \label{eq:HamiltonEqSY}
\end{align}
There exist many Hamiltonian flow lines in phase space $(\phi_c,\Pi_\Delta)$ that represent the solution of Hamilton's equations as shown in FIG.\ref{fig:HamiltonFlow}. Each flow line can be specified by the value of Hamiltonian since it is conserved on the given flow line. 
In the FIG.\ref{fig:HamiltonFlow}, there are two important flow lines corresponding to  $H(\phi_c, \Pi_{\Delta})=0$. 
Since the Hamiltonian is the quadratic polynomial of $\Pi_{\Delta}$, there are two $\Pi_{\Delta}$ satisfying $H(\phi_c, \Pi_{\Delta})=0$:
\begin{align}
\Pi_{\Delta} = 0, \, -\frac{8\pi^2}{3H^{4}}V'\ . \label{eq:MomentumZero}
\end{align}
In the case of $\Pi_{\Delta}=0 $, the equations \eqref{eq:HamiltonEqSY} reduce to
\begin{align}
\dot{\phi}_c &= -\frac{V'(\phi_c)}{3H} \ . \label{eq:HamiltonEqSYZero}
\end{align}
Similarly, in the case of $\Pi_{\Delta} =  -\frac{8\pi^2}{3H^{4}}V'(\phi_c)$, the equations \eqref{eq:HamiltonEqSY} reduce to
\begin{align}
\dot{\phi}_c = \frac{V'(\phi_c)}{3H}  \ .
\label{eq:HamiltonEqSYNonZero}
\end{align}
These two flow lines have three intersections  which correspond to the false vacuum, the top of potential hill and the true vacuum from left to right.
On the flow line with $\Pi_{\Delta}=0$, the false and true vacuum are stable and the top of potential is unstable.
While on the flow line with $\Pi_{\Delta} \neq 0$, the stability is reversed. The different signs of the potential forces account for the reverse of the stability at the intersections.
We see that the tunneling configuration exists; starting from the false vacuum $(\phi_c, \Pi_{\Delta})=(-\alpha, 0)$, going to the top of the potential $(\phi_c, \Pi_{\Delta})=(-\beta\alpha, 0)$ through the flow line with $\Pi_{\Delta} \neq 0$ and finally going to the true vacuum through the flow line with  $\Pi_{\Delta}=0$.
We numerically solve the second equation with the initial conditions $\phi_c(t')=-\alpha$ and switch to the first equation when the field reaches the top of potential $\phi_c(t)=-\beta\alpha$.
Thus, we obtain the tunneling configuration in FIG.\ref{fig:HomoBounce}. 

Now, we evaluate the transition probability \eqref{eq:ConditionalSY} by substituting the above tunneling configurations. The action in \eqref{eq:ConditionalSY} can be calculated without the concrete solutions for $\phi_c$;
\begin{align}
I
= \int_{t'}^{t} \mathrm{d}t \left[ \Pi_{\Delta}\dot{\phi}_c - H(\phi_c, \Pi_{\Delta})\right]
= -\frac{8\pi^{2}}{3H^{4}} \int_{t'}^{t_{*}} \mathrm{d}t \dot{\phi}_c V'(\phi_c)
= -\frac{8\pi^{2}}{3H^{4}} \Delta V, \label{eq:ActionHMSY}
\end{align}
where $t_{*}$ is the time that $\phi_c$ reaches the top of potential and $\Delta V$ is the difference of energy density between the top of the potential and the false vacuum. The trajectory from the top of the potential to the true vacuum does not contribute to the transition probability because $H(\phi_c,\Pi_{\Delta})=0$ and $\Pi_{\Delta}=0 $ on its trajectory. 
Thus, the transition probability of the tunneling can be evaluated as
\begin{equation}
 p(\alpha,t|-\alpha,t') \sim \exp\bigg(-\frac{8\pi^{2}}{3H^{4}}\Delta V\bigg). \label{eq:ConditionalHM}
\end{equation}
This shows a complete agreement with the result of the HM instanton on the fixed background \cite{Weinberg:2006pc}. 

Comments on this result are in order.
It is pointed out in \cite{Weinberg:2006pc} that when the potential $V(\phi)$ has several degenerate local maxima, all of them yield the same tunneling action in the Euclidean method, despite differing distances from the false vacuum. In our formalism, however, a transition to a distant maximum is described by a saddle-point solution that passes through all the intermediate maxima, and the transition probability is given by the product of the factors on the RHS of \eqref{eq:ConditionalHM} for each maximum. The aforementioned subtlety is absent in our formalism. 
%

Since the analysis in sec.~\ref{subsec:HM_onedim} is based on \eqref{eq:LangevinSY} describing the stochastic dynamics of the IR field $\phi_c$ at a single spatial point, eq.~\eqref{eq:ConditionalHM} calculates the tunneling probability at a single spatial point, not at the whole universe.\footnote{This is consistent with the observation made in \cite{Brown:2007sd} that HM solution corresponds to the transition over a region of a Hubble horizon volume.} Hence, physically speaking, \eqref{eq:ConditionalHM} provides the tunneling probability of a coarse-grained patch with a physical radius $(\varepsilon H)^{-1}$. This point becomes manifest also in sec.~\ref{subsec:HM_full}.

\subsection{The case for the full Langevin equation}\label{subsec:HM_full}
Next, we evaluate the tunneling rate based on the formula \eqref{eq:ConditionalFull}. An advantage of \eqref{eq:ConditionalFull} is that one can discuss various non-trivial dynamics in the whole universe covered by the flat chart in principle.

In the previous section, we focused on the dynamics at a single spatial point and found the HM tunneling process as a non-trivial solution of the Hamilton's equations. In this section, we discuss the corresponding process in the global picture based on \eqref{eq:ConditionalFull} by identifying a non-trivial configuration that satisfies the Hamilton's equations in a good approximation. We then  reproduce the result \eqref{eq:ActionHMSY}. We also estimate the characteristic time scale of the tunneling. In sec.~\ref{sec:remark}, we remark some technical complications that arise in the global picture.

We start with simplifying the expression \eqref{eq:ConditionalFull}. As discussed in section~\ref{sec:Stochastic_approach},
$G^{\phi\phi}$ becomes dominant and the others can be neglected for $\varepsilon \ll 1$ in \eqref{eq:def_Hamiltonian}. Under this approximation, the exponent of the integrand of \eqref{eq:ConditionalFull} is linear in $\phi_\Delta$. We can then perform the path integral $\int\mathcal{D}\phi_\Delta$ in \eqref{eq:ConditionalFull}, yielding the product of delta functions $\prod_{\bm x}\delta\bigl(\dot\Pi_c-a\nabla^2\phi_c+a^3V'(\phi_c)\bigr)$. These delta functions give an equation of motion for $\Pi_c$ and eliminate the path integral $\int\mathcal{D}\Pi_c$. We solve this equation as $\Pi_c\simeq -a^3V'(\phi_c)/3H$, ignoring the spatial gradient and the time variation of $\phi_c$. Furthermore, we change the integration variable as $\Pi_\Delta\to-i\Pi_\Delta$ in \eqref{eq:ConditionalFull}. Consequently, \eqref{eq:ConditionalFull} reduces to
\begin{equation}
    p(\phi_c(\bm{x}),t |\phi_c'(\bm{x}), t')
    \simeq
    \int_{\phi_c(t',\bm{x})=\phi_c'(\bm{x})}^{\phi_c(t,\bm{x})=\phi_c(\bm{x})} \mathcal{D}(\phi_c, \Pi_{\Delta})  
    \exp\left[ 
        \int \mathrm{d}^{4}x 
        \left(
            \Pi_{\Delta}\dot{\phi}_c - H_\text{HM}(\phi_c, \Pi_{\Delta})
        \right)
        \right] 
        \,,\label{eq:HMfull}
\end{equation}
where the Hamiltonian $H_\text{HM}$ is defined as 
\begin{align}
    H_\text{HM}(\phi_c,\Pi_\Delta)
   & \coloneqq
    -\frac{1}{3H}V'(\phi_c)\Pi_\Delta
    -\frac{H^{3}}{8\pi^{2}}
    \Pi_\Delta\overline\Pi_\Delta
    \,,\label{eq:HMHamiltonian}
\end{align}
and $\overline \Pi_\Delta$ is given by  
\begin{align}
    \overline\Pi_\Delta(t,\bm x)
    &\coloneqq
    \int \mathrm{d}^{3}x' \, 
    j_0\left(k_c(t)|\bm x - \bm x'|\right)
    \Pi_{\Delta}(t,\bm x')
    =
    \int\frac{\mathrm{d}^3 k}{(2\pi)^3}\,
    \frac{4\pi^2}{H^3}\frac{H^2}{2k^3}
    \delta(t-t_k) \Pi_\Delta(t,\bm k) e^{i\bm k\cdot\bm x}
    \,.\label{def:pibar}
\end{align}
Here we define $t_k$ by the condition $k_c(t_k)=k$. The Hamilton's equations are
\begin{subequations}
\label{eq:HMeom}
\begin{align}
    &\dot{\phi}_c 
    = 
    -\frac{1}{3H}V'(\phi_c)
    - \frac{H^3}{4\pi^2} \overline\Pi_\Delta
    \,,\label{eq:HMeom_a}\\
    &\dot{\Pi}_\Delta 
    = 
    \frac{1}{3H}V''(\phi_c)\Pi_\Delta
    \,.\label{eq:HMeom_b}
\end{align}
\end{subequations}
Here, the term $\dot\Pi_\Delta(t,\bm x)$ on the LHS of \eqref{eq:HMeom_b} is defined as the Fourier transform of $\dot\Pi_\Delta(t,\bm k)$ i.e., $\dot\Pi_\Delta(t,\bm x)\coloneqq(2\pi)^{-3}\int\mathrm{d}^3 k\,\dot\Pi_\Delta(t,\bm k) e^{i\bm k\cdot \bm x}$. The same rule applies to other IR variables in the real space as in eq.~\eqref{real_timederiv}. Intuitively, this is because the stochastic formalism is formulated in the momentum space and its Fourier transform provides the stochastic formalism in the real-space: see appendix~\ref{sec:in-in} for more details. 

Eqs.~\eqref{eq:HMeom} admit a trivial configuration $\Pi_\Delta(t,\bm k)=0$ under which we have $H_\text{HM}=0$ and \eqref{eq:HMeom_a} becomes $\dot\phi_c=-V'(\phi_c)/(3H)$ describing the standard classical time evolution. Now we are interested in the non-trivial configuration which satisfies \eqref{eq:HMeom}. In the previous section, we focus on the non-trivial dynamics at the single spatial point $\bm x = \bm x_0$ and it is found that the tunneling configuration is obtained by $\Pi_\Delta(t,\bm x_0) = -\frac{8\pi^2}{3H^4} V'(\phi_c(t,\bm x_0))$. This configuration can be naturally extended to the one in the global picture by considering the following configuration 
\begin{align}
    \Pi_\Delta(t,\bm k)
    =
    -\frac{8\pi^2}{3H^4}V'(\phi_c(t,\bm x_0))
    e^{-i\bm k\cdot \bm x_0}
    \,.\label{configmom_HM}
\end{align}
From now on, we show that this configuration satisfies a set of Hamilton equations \eqref{eq:HMeom} in a good approximation. For this configuration, we have
\begin{subequations}
\label{config_HM}
\begin{align}
    &\Pi_\Delta(t,\bm x)
    =
    -\frac{8\pi^2}{3H^4}V'(\phi_c(t,\bm x_0)) 
    f(r;t)
    \,,\label{eq:pi}
    \quad
    \\
    &\dot\Pi_\Delta(t,\bm x)
    =
    -\frac{8\pi^2}{3H^4}V''(\phi_c(t,\bm x_0))\dot\phi_c(t,\bm x_0) 
    f(r;t)
    \,,\label{eq:pidot}
    \\
    &\overline\Pi_\Delta(t,\bm x)
    =
    -\frac{8\pi^2}{3H^4}V'(\phi_c(t,\bm x_0)) 
    j_0\left(k_c(t)r\right)
    \,,\label{eq:pibar}
\end{align}
\end{subequations}
where $r\coloneqq |\bm x -\bm x_0|$ and $f(r;t)\coloneqq (2\pi)^{-3}\int\mathrm{d}^3k\,\theta(k_c(t)-k)\,e^{i\bm k\cdot(\bm x-\bm x_0)}$. The functions $j_0(k_c(t)r)$ and $f(r;t)$ rapidly oscillate for $r\gg k_c^{-1}(t)$. Hence, they exponentially decay to zero after they are smeared over the Hubble time: for instance, $j_0$ and $f$ becomes the smooth function after the suitable coarse-graining as~\footnote{
Note that this behavior may motivate one to replace the term $j_0(k_c(t)r)$ in \eqref{eq:HMHamiltonian} by the step function 
\begin{align}
    j_0(k_c(t)r)
    \rightarrow
    \theta(1-k_c(t)r)
    \,.
\end{align}
This is the approximation adopted in \cite{Starobinsky:1994bd}.
} 
\begin{align}
    j_0(k_c(t)r)
    &\xrightarrow{\text{time c.g.}}
    W(r;t)
    \coloneqq 
    \int\mathrm{d}t' w(t',t) 
    j_0\left(k_c(t')r\right)
    \approx
    \begin{cases}
        1 & (r\ll k_c^{-1}(t)) \\
        0 \quad 
        & (r \gg k_c^{-1}(t))
    \end{cases}
    \no
    \,,\\
    f(r;t) 
    &\xrightarrow{\text{time c.g.}}
    W_f(r;t)\coloneqq \int\mathrm{d}t'\,w(t',t) f(r;t')
    \approx
    \begin{cases}
        k_c^3(t)/(6\pi^2) & (r\ll k_c^{-1}(t)) \\
        0 \quad 
        & (r \gg k_c^{-1}(t))
    \end{cases}
    \label{spatialwindow_2}
    \,.
\end{align}
Here, we perform the smearing by using a window function $w(t',t)$ which is approximately constant in the domain $|t-t'|\ll H^{-1}$ while it exponentially decays to zero at $|t-t'|\gg H^{-1}$. We also impose the normalization condition $\int \mathrm{d} t' w(t',t) =1$. We can choose $w(t',t)=H\pi^{-1/2}\exp[-H^2(t-t')^2]$ for example. 
Note that the property \eqref{spatialwindow_2} is based on the behavior $j_0(k_cr)\simeq 1$ and $f(r;t)\simeq k_c^3/(6\pi^2)$ for $r\ll k_c^{-1}$ and the rapid oscillations of $j_0(k_cr)$ and $f_0(r;t)$ for $r\gg k_c^{-1}$. Hence, the property \eqref{spatialwindow_2} will be robust against the choice of $w(t',t)$. 

Therefore, substituting Eqs.~\eqref{eq:pibar} and \eqref{spatialwindow_2} into Eq.~\eqref{eq:HMeom_a} and performing the smearing in time, we obtain
\begin{subequations}
\begin{align}
    \dot\phi_c(t,\bm x)
    =
    &-\frac{1}{3H}
    \left[
    V'(\phi_c(t,\bm x))
    -2 V'(\phi_c(t,\bm x_0)) j_0\left(k_c(t)r\right)
    \right]
    \label{phicdot_exact}\\
    &\xrightarrow{\text{time c.g.}}
    \begin{cases}
        -V'(\phi_c(t,\bm x))/(3H) & (r \gg k_c^{-1}(t)) \\
        V'(\phi_c(t,\bm x_0))/(3H) & (r \ll k_c^{-1}(t))
    \end{cases}
    \,.\label{phicdot_smear}
\end{align}
\end{subequations}
This exhibits the non-trivial dynamics only in the domain $r\lesssim k_c^{-1}$, i.e., a coarse-grained patch centered at a point $\bm x_0$ with a physical radius $(\varepsilon H)^{-1}$. Here, we used $\phi_c(t,\bm x)\simeq \phi_c(t,\bm x_0)$ for $r\ll k_c^{-1}(t)$. In the second line, it is assumed that $V'(\phi_c(t,\bm x_0))$ does not cancel the rapid oscillations of $j_0(k_c(t)r)$ so that their product $V'j_0$ decays exponentially at $r\gg k_c^{-1}$ after the suitable smearing. This assumption will be valid thanks to the approximate constancy of super-Horizon fluctuations $\phi_c(t,\bm k)$ over the Hubble time: $\phi_c(t,\bm k)\simeq \phi_c(t',\bm k )$ for $|t-t'|\lesssim H^{-1}$. Adopting the same assumption, from \eqref{eq:pi} and \eqref{eq:pidot} we also obtain the smeared expressions of $\Pi_\Delta$ and $\dot\Pi_\Delta$ which turn out to satisfy \eqref{eq:HMeom_b} under the condition \eqref{eq:HMeom_a}.
Hence, we conclude that \eqref{config_HM} describes a non-trivial saddle-point solution in a good approximation. Furthermore, after the smearing we can find $H_\text{HM}\approx 0$ for this configuration.

Now we use the configuration \eqref{configmom_HM} or \eqref{config_HM} to obtain the tunneling configuration shown in FIG.\ref{fig:HomoBounce} again. We can calculate the action for this tunneling configuration similarly to \eqref{eq:ActionHMSY}. By substituting the exact expressions \eqref{config_HM} and \eqref{phicdot_exact}, we obtain~\footnote{Note that we can also estimate the action $I$ by performing the coarse-graining in time suitably and substituting the smeared expression of $\Pi_\Delta$ into the action:  
\begin{align}
    I
    &\simeq 
    -\frac{8\pi^2}{3H^4}\int_{t'}^{t_{*}} \mathrm{d}t\,
    V'(\phi_c(t,\bm x_0))
    \int\mathrm{d}^3x\, \dot{\phi}_c(t,\bm x) 
    W_f(r;t)
    \simeq 
    -\frac{8\pi^2}{3H^4}\int_{t'}^{t_{*}}\mathrm{d}t\,
    \dot\phi_c(t,\bm x_0)V'(\phi_c(t,\bm x_0))
    = -\frac{8\pi^{2}}{3H^{4}} \Delta V\,,
    \no
\end{align}
reproducing the previous result \eqref{eq:ActionHMSY}. In the first approximate equality, we used $H_\text{HM}=0$ which is well satisfied after the smearing. In the second one, we use $\phi_c(t,\bm x)\simeq \phi_c(t,\bm x_0)$ for $r\ll k_c^{-1}(t)$ and the condition $\int\mathrm{d}^3x\,W_f(r;t)=1$ which follows from the definition of $W_f$. This estimate may be more analogous to the calculation of the action in sec.~\ref{subsec:HM_onedim} since the approximate locality of the dynamics is restored after the smearing.} 
\begin{align}
    I
    &=
     -\frac{H^3}{8\pi^2}\int_{t'}^{t_{*}} \mathrm{d}t\,
    \int\mathrm{d}^3x\, \Pi_\Delta(t,\bm x)\overline\Pi_\Delta(t,\bm x)
    \no\\
    &= 
    -\frac{8\pi^2}{3H^4}\int_{t'}^{t_{*}} \mathrm{d}t\,
    V'(\phi_c(t,\bm x_0)) \frac{V'(\phi_c(t,\bm x_0))}{3H}
    \int\mathrm{d}^3x\, j_0\left(k_c(t)r\right) f(r;t)
    \no\\ 
    &=
    -\frac{8\pi^2}{3H^4}\int_{t'}^{t_{*}}\mathrm{d}t\,
    \dot\phi_c(t,\bm x_0)V'(\phi_c(t,\bm x_0))
    = -\frac{8\pi^{2}}{3H^{4}} \Delta V\,.
    \label{eq:HM_est}
\end{align}
In the first line, we used \eqref{eq:HMeom_a} to eliminate $\dot\phi_c$ from the integrand. In the second line, we substituted the concrete configurations \eqref{config_HM}. In the third line, we used \eqref{phicdot_exact} and performed the spatial integral as $\int\mathrm{d}^3x\, j_0\left(k_c(t)r\right) f(r;t)=1$. Thus, we reproduce the previous result \eqref{eq:ActionHMSY}. Our analysis confirms that \eqref{eq:HM_est} gives the tunneling probability of a coarse-grained patch with a physical radius $(\varepsilon H)^{-1}$.

It is worth mentioning the time scale of the HM tunneling. We may naively calculate the typical time scale of the HM tunneling as $t_*-t'=\int^{\phi_\text{top}}_{\phi_\text{false}}\mathrm{d}\phi_c\,\frac{3H}{V'(\phi_c)}$. Here, the space-time argument of the field is suppressed. However, this integral is in general divergent since $V'=0$ at the both ends of the integral. In the vicinity of the domain where $V'=0$, we expect that the quantum fluctuations of the field would play an important role. Since the typical size of quantum fluctuations accumulated over the Hubble time is $H/(2\pi)$, we may define the regulated quantity $t_\text{HM}$ as  
\begin{align}
    t_\text{HM}
    \coloneqq
    \int^{\phi_\text{top}-\frac{H}{2\pi}}_{\phi_\text{false}+\frac{H}{2\pi}}\mathrm{d}\phi_c
    \,
    \frac{3H}{V'(\phi_c)}
    \,,\label{timescale}
\end{align}
and we expect that $t_\text{HM}$ would correctly characterize the typical time scale of the HM tunneling. In particular, for our potential \eqref{eq:potential} we have 
\begin{align}
    t_\text{HM}
    &=
    \frac{3H}{2g^2 \alpha ^2 \left(1+\beta\right)}\left[\left(\frac{3+\beta}{1-\beta}\right) \log \left|1-\frac{1-\beta}{H/2\pi\alpha}\right| - \log \left|\frac{1-\frac{2}{H/2\pi\alpha}}{1+\frac{1+\beta}{H/2\pi\alpha}}\right|\right]
    \sim
    \frac{1}{H}\cdot\mathcal{O}\left((H/g\alpha)^2\right)
    \,.\label{timescale_2}
\end{align}
 Since we have $(H/g\alpha)^2\sim (H^2/V'')_{\phi_\text{false}\lesssim \phi_c\lesssim \phi_\text{top}}\gg1$ for the shallow potential, we conclude that the time scale of HM tunneling is much longer than the Hubble time. 

\subsubsection{Remark} \label{sec:remark}
So far it is found that, starting from \eqref{eq:ConditionalFull}, the configuration \eqref{configmom_HM} approximately solves the Hamilton's equations \eqref{eq:HMeom} and reproduce the previous result \eqref{eq:ActionHMSY} when it is substituted into the action. There is actually an important reason behind why we should choose the configuration \eqref{configmom_HM} to evaluate the HM tunneling process. 

To see this, it is important to realize an important difference between the current Hamilton's equations \eqref{eq:HMeom} and the previous one \eqref{eq:HamiltonEqSY}; the equation for $\phi_c$ now contains $\overline{\Pi}_\Delta$ rather than $\Pi_\Delta$. Due to this difference, the structures of Hamilton's equations \eqref{eq:HMeom} are understood as follows;
\begin{itemize}
    \item the equation for $\phi_c$ \eqref{eq:HMeom_a} is determined once $\overline\Pi_\Delta$ is specified,
    \item $\overline{\Pi}_\Delta(t,\bm x)$ is understood as initial conditions for \eqref{eq:HMeom_b} since $\overline{\Pi}_\Delta(t,\bm x)$ contains only the boundary modes satisfying $k=k_c(t)$. Hence, $\Pi_\Delta(t,\bm x)$ is obtained by solving \eqref{eq:HMeom_b} for given $\overline{\Pi}_\Delta$ under the condition \eqref{eq:HMeom_a}. 
\end{itemize}
On top of that, substituting the Hamilton's equation for $\phi_c$ into the action, we have the first line of \eqref{eq:HM_est} which can be rewritten as
\begin{align}
    I
    =
    -\frac{1}{2}\int^{t_*}_{t'}\mathrm{d}t\,
    \int\frac{\mathrm{d}^3k}{(2\pi)^3}\,
    \Pi_\Delta(t,\bm k)\Pi_\Delta(t,-\bm k)
    \frac{H^2}{2k^3}\delta(t-t_k)
    \,.\label{eq:HM_est_2}
\end{align}
The RHS depends only on the value of $\Pi_\Delta(t,\bm k)|_{k=k_c(t)}$.
Hence, we can evaluate the tunneling action once $\overline{\Pi}_\Delta$ is specified, provided that \eqref{eq:HM_est_2} or equivalently \eqref{eq:ConditionalFull} is reliable. 

We claim that \eqref{eq:ConditionalFull} will be valid when $\overline{\Pi}_\Delta$ is chosen so that $\Pi_\Delta(t>t_k,\bm k)$ becomes a smooth function in time. Otherwise, it will not be justified to ignore the higher-order corrections $\delta H_\text{higher}$ which are neglected in \eqref{eq:ConditionalFull}. Intuitively, this states that perturbation theory tend to be broken down around exotic configurations. 
When $\Pi_\Delta$ is a smooth function, we can relate $\overline{\Pi}_\Delta$ to $\Pi_\Delta$ after the suitable
smearing in time:
\begin{align}
    &\overline{\Pi}_\Delta(t,\bm x)
    \xrightarrow{\text{time c.g.}}
    \int\mathrm{d}^3 x'
    \Pi_\Delta (t,\bm x') W(|\bm x-\bm x'|;t)
    \sim 
    \mathcal{V}\Pi_\Delta(t,\bm x)
    \,.    \label{eq:prac_repl}
\end{align}
Here, we also used $\Pi_\Delta(t,\bm x)\sim \Pi_\Delta(t,\bm x')$ for $|\bm x-\bm x'|\lesssim k_c^{-1}$ and defined $\mathcal{V}\coloneqq(4\pi/3)k_c^{-3}(t)\sim\int\mathrm{d}^3x'\,W(|\bm x-\bm x'|;t)$.
We then require that the Hamilton's equation for $\Pi_\Delta$ \eqref{eq:HMeom_b} is satisfied in a good approximation to ensure the validity of \eqref{eq:ConditionalFull}. This constrains the choice of $\overline\Pi_\Delta$. Indeed, the configuration \eqref{configmom_HM} approximately solves \eqref{eq:HMeom_b} and satisfies the smoothness and henceforth \eqref{eq:prac_repl} after the smearing. This would be the reason why our analysis with \eqref{configmom_HM} based on \eqref{eq:ConditionalFull} can reproduce the correct result. 

\subsection{On the \texorpdfstring{$\varepsilon$}{e}-independence of the Hawking-Moss tunneling}
In sections~\ref{subsec:HM_onedim} and \ref{subsec:HM_full}, the tunneling probability of a coarse-grained patch with a physical radius $(\varepsilon H)^{-1}$ is calculated. Our results coincide with the HM tunneling and are independent of $\varepsilon$. 

The $\varepsilon$-independence of the results would be a consequence of the scale-independence of the dynamics of light scalar fields at super-horizon scales. To see this, it is useful to notice that a coarse-grained patch of physical radius $(\varepsilon_1 H)^{-1}$ at a time $t=T$ expands to a patch of larger physical radius $(\varepsilon_2H)^{-1}$ at a later time $t=T+\delta t>T$ with $\varepsilon_2=\varepsilon_1\exp[-H\delta t]<\varepsilon_1\ll1$.  
The value of IR field does not evolve from $t=T$ to $T+\delta t$ because the fluctuations of light scalar field at super-horizon scales are approximately time-independent,
\begin{align}
    \int\frac{\mathrm d^3k}{(2\pi)^3}\,\phi_c(t,\bm k)  e^{i\bm k.\bm x}\theta(k_c(t)-k)
    \simeq
    \int\frac{\mathrm d^3k}{(2\pi)^3}\,\phi_c(t+\delta t,\bm k)  e^{i\bm k.\bm x}\theta(k_c(t)-k)
    \,,\label{eq:fieldmap1}
\end{align}
unless $\delta t$ is much longer than $H^{-1}$. Hence, we can relate the coarse-grained dynamics of different $\varepsilon\ll1$ by considering the time shift $t\to t+\delta t$ with the value of IR field being kept fixed. This means that the tunneling probability from $\phi=\phi_\text{false}$ to $\phi_\text{true}$ of a patch with physical radius $(\varepsilon H)^{-1}$ should be independent of $\varepsilon$.

\section{Coleman-de Luccia tunneling}\label{sec:CDL}
\begin{figure}[btp]
\centering
\includegraphics[width=100mm]{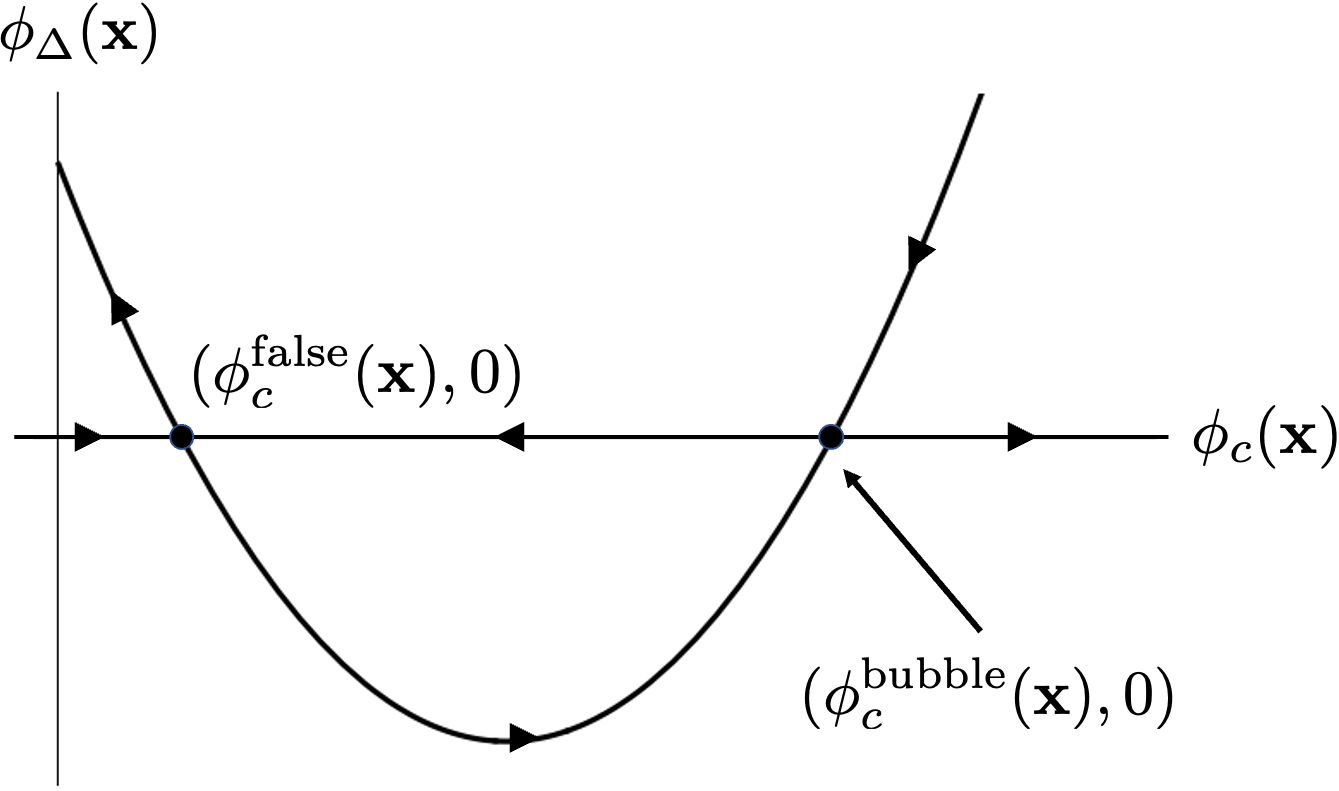}
\caption{A schematic picture of Hamiltonian flows in the "field" phase space. We expect that there is a flow line  which starts from the false vacuum (left intersection) and reaches the bubble configuration (right intersection) through a non-zero $\phi_{\Delta}(\bm{x})$ path.}
\label{fig:configuration_flow}
\end{figure}

In the previous section, the result of the HM instanton is reproduced by using \eqref{eq:ConditionalFull}.
We expect that the formula \eqref{eq:ConditionalFull} can describe not only the HM tunneling but also the CDL tunneling.
In other words, we expect that there will be a flow line starting from the false vacuum to the bubble configurations as illustrated in FIG. \ref{fig:configuration_flow}. 
In this section, we concretely show an interesting configuration by following the prescription in the previous section.

\subsection{Coleman-de Luccia bubble solution}\label{sec:bubble_sol}
In Euclidean method, the CDL instanton is dominant rather than the HM instanton when the potential barrier is steep \cite{Jensen:1983ac,Jensen:1988zx,Hackworth:2004xb,Batra:2006rz}.
By taking $\varepsilon \gg 1$, we can study nonperturbative physics at sub-horizon scales such as the formation of bubble. In this case, $G^{\Pi\Pi}$ becomes dominant and the other noises can be ignored in \eqref{eq:def_Hamiltonian}. Under this approximation, the exponent of the integrand of \eqref{eq:ConditionalFull} is linear in $\Pi_\Delta$. We can then perform the path integral $\int\mathcal{D}\Pi_\Delta$ in \eqref{eq:ConditionalFull}, yielding the product of delta functions $\prod_{\bm x}\delta\bigl(\dot\phi_c-a^{-3}\Pi_c\bigr)$. These delta functions eliminate the path integral $\int\mathcal{D}\Pi_c$.\footnote{This is the reason why we illustrate flow lines in the $(\phi_c,\phi_\Delta)$-plane in FIG.~\ref{fig:configuration_flow}.} Furthermore, we change the integration variable as $\phi_\Delta\to-i\phi_\Delta$ in \eqref{eq:ConditionalFull}, leading to 
\begin{equation}
    p
    \simeq
    \int
    \mathcal{D}(\phi_c, \phi_{\Delta}) 
    \exp\left[ 
        \int \mathrm{d}^{4}x \,
            \left(
                 a^3\dot \phi_{\Delta}\dot{\phi}_c 
                - H_\text{CDL}(\phi_c, \phi_{\Delta})
            \right)
        \right] 
        \label{eq:CDLfull}
\end{equation}
with appropriate boundary conditions; we consider bubble configurations as boundary conditions later. Here, the Hamiltonian $H_\text{CDL}$ is defined by 
\begin{equation}
    H_\text{CDL}(\phi_c,\phi_\Delta)
    \coloneqq
    -\left(
        a\nabla^2\phi_c-a^3V'(\phi_c)
    \right)\phi_\Delta
    -
    \frac{a^3}{2}\phi_\Delta(x)\overline\phi_\Delta(x)
    \,,   
    \label{eq:CDLHamiltonian}
\end{equation}
with
\begin{align}
    \overline{\phi}_\Delta(t,\bm x)
    \coloneqq
    \frac{Hk_c^4(t)}{4\pi^2a(t)}\int\mathrm{d}^3x'\,
    j_0\left(k_c(t)|\bm x - \bm x'|\right)
    \phi_\Delta(t,\bm x')
    \,.
\end{align}
We would like to find the appropriate configuration $\overline{\phi}_\Delta$ which describes the tunneling process. Following the choice made in sec.~\ref{sec:HM}, let us {\it suppose} that the appropriate choice is given by the configuration for which the Hamiltonian \eqref{eq:CDLHamiltonian} vanishes:
\begin{equation}
    \overline\phi_{\Delta}(x) 
    = 0
    \,,\,
    -2(a^{-2}\nabla^{2}\phi_c - V'(\phi_c))
    \,.
    \label{eq:MomentumZeroCDL_bar}
\end{equation}
Note that this condition is imposed up to the suitable smearing in time because $\overline{\phi}_\Delta(t,\bm x)$ includes only the boundary Fourier mode $k=k_c(t)$. As discussed in sec.~\ref{sec:remark}, $\phi_\Delta$ becomes sufficiently smooth for the appropriate $\overline{\phi}_\Delta$, and hence we also suppose 
\begin{align}
    \overline\phi_\Delta
    \sim
    \frac{H^{2}\varepsilon}{3\pi}
    \phi_{\Delta}(x)
\label{eq:noise_CDL}
\end{align}
after the suitable averaging over the Hubble time, similarly to \eqref{eq:prac_repl}. We use the relation \eqref{eq:noise_CDL} to evaluate the action later. This would work at least for the purpose of estimating the action.

In principle, we do not need to use the estimate \eqref{eq:noise_CDL} in evaluating the action since it is determined once $\overline{\phi}_\Delta$ is specified based on the analogous logic discussed around \eqref{eq:HM_est_2}. For this purpose, we need to specify the configurations $\{\phi_\Delta(t,\bm k)|_{k=k_c(t)}\}$ in momentum space which result in the sufficiently smooth $\phi_\Delta$ that solves the Hamilton equations. It is not easy to find such appropriate configurations precisely, however. Hence, we leave more careful analysis on the choice of $\overline{\phi}_\Delta$ for future work. In this study, instead, we focus on how we can proceed the analysis for given configuration \eqref{eq:MomentumZeroCDL_bar} with adopting the estimate \eqref{eq:noise_CDL} and how the bubble nucleation process could be described in the stochastic approach.

Using the relation \eqref{eq:noise_CDL}, the Hamiltonian \eqref{eq:CDLHamiltonian} and the configuration \eqref{eq:MomentumZeroCDL_bar} become 
\begin{align}
    &H_\text{CDL}
    \sim
    -
    \left(
        a\nabla^2\phi_c-a^3V'(\phi_c)
    \right)\phi_\Delta
    -
    \frac{H^{2}\varepsilon}{6\pi}a^3\phi_{\Delta}^{2}
    \,, \label{eq:HamiltonianInhomo}
    \\
    &\phi_{\Delta}(x) 
    = 0
    \,,\,
    -\frac{6\pi}{H^2\varepsilon}(a^{-2}\nabla^{2}\phi_c - V'(\phi_c))
    \,.
    \label{eq:MomentumZeroCDL}
\end{align}
The Hamilton's equations under the constraints \eqref{eq:MomentumZeroCDL} give the following equation of motions for $\phi_c$:
\begin{align}
\ddot{\phi}_c + 3H\dot{\phi}_c&= -(V'(\phi_c) -a^{-2}\nabla^{2}\phi_c) , \quad (\phi_{\Delta}=0), \notag\\
\ddot{\phi}_c + 3H\dot{\phi}_c &= V'(\phi_c)-a^{-2}\nabla^{2}\phi_c , \qquad\,\,\, (\phi_{\Delta} = -\frac{6\pi}{H^{2}\varepsilon}(a^{-2}\nabla^{2}\phi_c - V'(\phi_c))). \label{eq:HamiltonEqInhomoZero}
\end{align}
Similar to the previous case \eqref{eq:HamiltonEqSYZero}, the signs of the potential term and the gradient term are flipped in the equation of motion with the non-trivial $\phi_\Delta\neq0$. Hence, we can expect that the tunneling process is realized.
From \eqref{eq:CDLfull}, we find that only the solutions satisfying $\phi_{\Delta}\neq 0$ contribute to the action. Then we focus on the second equation in \eqref{eq:HamiltonEqInhomoZero}.

Interestingly, this equation is the classical equation of motion in the Euclidean anti-de Sitter (AdS) space which is defined by the embedding equation
\begin{align}
-X_{0}^{2} + X_{1}^{2} + X_{2}^{2} + X_{3}^{2} +X_{4}^{2} = -H^{-2}
\end{align}
in five-dimensional Minkowski spacetime
\begin{align}
\dm s^{2} = -\dm X_{0}^{2} + \dm X_{1}^{2} + \dm X_{2}^{2} + \dm X_{3}^{2} +\dm X_{4}^{2}    \, .
\end{align}
In fact, the d'Alembert operator $\partial_{t}^{2} + 3H\partial_{t} + a^{-2}\nabla^{2}$ can be obtained
from the following induced metrics;
\begin{align}
\dm s^{2} 
= \mathrm{d}t^{2}+e^{2Ht}\dm\bm{x}^{2}
= \frac{1}{H^{2}\eta^{2}}(\dm\eta^{2}+\dm\bm{x}^{2}) 
= H^{-2}(\dm\rho^{2} + \sinh^{2}\rho\, \dm\Omega^{2})
\ , \label{eq:MetricsEAdS}
\end{align}
where $\dm\Omega^{2}:=\dm\theta_{1}^{2} + \sin^{2}\theta_{1}\dm\theta_{2}^{2} + \sin^{2}\theta_{1}\sin^{2}\theta_{2}\dm\theta_{3}^{2}$ and $0<\rho<\infty$, $0\leq \theta_{1}\leq \pi$, $0\leq \theta_{2}\leq \pi$ and $0\leq \theta_{3} \leq 2\pi$.
The second metric is that in the Poincar\'{e} coordinates and the third one  is that in the global coordinates.
The explicit coordinate transformations are given by
\begin{align}
X_{0} &= \frac{1}{2}\frac{\eta^{2} + \bm{x}^{2} + H^{-2}}{-\eta} =H^{-1}\cosh\rho, \notag\\
X_{1} &= \frac{x_{1}}{-H\eta} = H^{-1}\sinh\rho \sin\theta_{1} \cos\theta_{2}, \notag\\
X_{2} &= \frac{x_{2}}{-H\eta} = H^{-1}\sinh\rho \sin\theta_{1} \sin\theta_{2} \cos\theta_{3}, \notag\\
X_{3} &= \frac{x_{3}}{-H\eta} = H^{-1}\sinh\rho \sin\theta_{1} \sin\theta_{2} \sin\theta_{3}, \notag\\
X_{4} &= \frac{1}{2}\frac{\eta^{2}+\bm{x}^{2}-H^{-2}}{-\eta}
=H^{-1}\sinh\rho \cos\theta_{1}.
\label{eq:Coordinate}
\end{align}
For later convenience, we derive the relation between $(\eta, r\coloneqq |\bm x|)$ and $(\rho,\sigma_1)$ as
\begin{equation}
    -H\eta = \frac{1}{\cosh\rho-\sinh\rho\cos\theta_1}\,,
    \qquad
    r= H^{-1}\frac{\sinh\rho\sin\theta_1}{\cosh\rho-\sinh\rho\cos\theta_1}
    \,.
\end{equation}
From this, we can check that a point $(\eta,r)=(-H^{-1},0)$ is mapped to $\rho=0$ for any $\theta_1\in[0,\pi]$. Except for this, there is one-to-one correspondence between points in the $(\eta,r)$-plane with $\eta\leq0$ and $r\geq0$ and points in the $(\rho,\theta_1)$-plane with $\rho>0$ and $0\leq \theta_1\leq\pi$. On a constant-$\eta$ slice in the $(\eta,r)$-plane, the value of $\rho$ for given spatial point $r$ is given by 
\begin{equation}
\rho = \text{arccosh}\bigg(\frac{1}{2}\frac{H^{2}(\eta^{2}+r^{2})+1}{-H\eta}\bigg).\label{eq:rho}
\end{equation}
This function increases monotonically in $r$ for given $\eta$ in the region $r>0$.

\begin{figure}[btp]
\centering
\includegraphics[width=120mm]{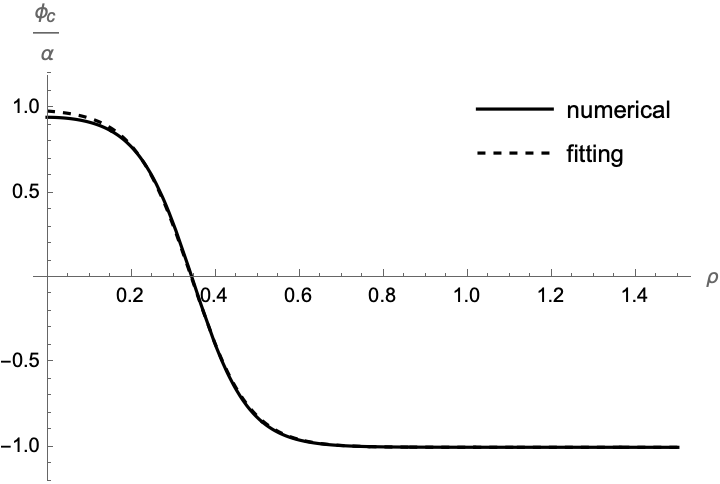}
\caption{The tunneling configuration for the inhomogeneous field is plotted. The solid curve is the numerical solution for $\beta=0.5$ and $\alpha g/H=\sqrt{140}$. 
We take the initial conditions as $\phi_c(10^{-15})/\alpha=1-4.98284\times10^{-2}$ and $\phi_c'(10^{-15})/\alpha=0$. 
The dashed curve is the fitting function \eqref{eq:Fitting} where $\mu=7.32515$ and $\bar{\rho}=0.34397$.}
\label{fig:InhomoBounce}
\end{figure}

From the global coordinate \eqref{eq:MetricsEAdS},
we see the Euclidean AdS spacetime has $O(4)$ symmetry.
Hence, we assume that the field $\phi_c$ depends only on $\rho$.
Thus the second equation of \eqref{eq:HamiltonEqInhomoZero} reads
\begin{align}
\frac{\dm^{2}\phi_c}{\dm\rho^{2}} + \frac{3}{\tanh\rho}\frac{\dm\phi_c}{\dm\rho} = \frac{V'(\phi_c)}{H^{2}}. \label{eq:HamiltonEqO4}
\end{align}
Note that this is the same equation utilized in the Euclidean method \cite{Coleman:1980aw,Rubakov:1999ir}.
Imposing the boundary conditions \cite{Coleman:1977py,Coleman:1980aw}
\begin{align}
\lim_{\rho \to \infty} \phi_c = -\alpha, \quad \frac{d\phi_c}{d\rho}\bigg|_{\rho=0} =0, \label{eq:BoundaryCondition}
\end{align}
we obtain the bubble solutions as is shown in FIG. \ref{fig:InhomoBounce}. Note that the latter condition ensures the continuity of $\partial_\eta\phi_c(\eta,r)$ and $\partial_\eta^2\phi_c(\eta,r)$ at $(H\eta,Hr)=(-1,0)$.

It is useful to fit the bubble configurations by the following fitting function (FIG. \ref{fig:InhomoBounce});
\begin{align}
\phi_c(\rho) = -\alpha\tanh \mu(\rho-\bar{\rho}),
\label{eq:Fitting}
\end{align}
where $\mu$ and $\bar{\rho}$ represent the thickness and the position of the bubble wall, respectively.
Note that, for $\beta \gtrsim 0.6$, the deviation from the true vacuum at the $\rho=0$ becomes significant \cite{Weinberg:2012pjx} and the fitting \eqref{eq:Fitting} becomes bad.

\begin{figure}[htbp]
\begin{tabular}{cc}

\begin{minipage}[t]{0.45\hsize}
\centering
\includegraphics[width=70mm]{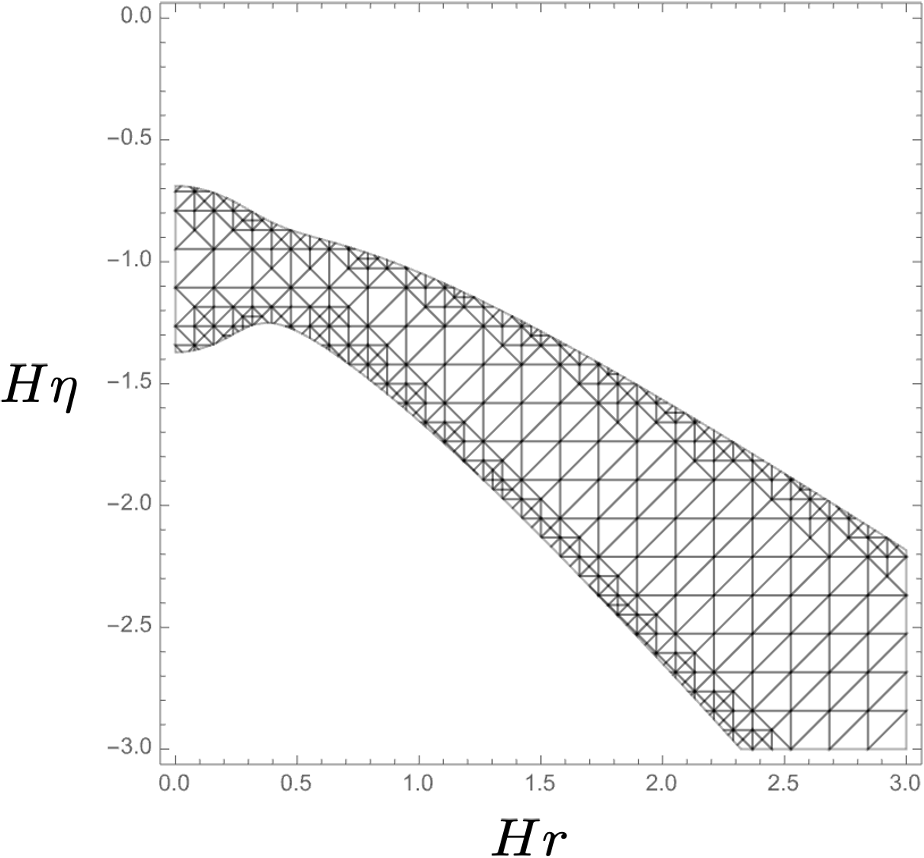}
\caption*{(a)}
\end{minipage} &

\begin{minipage}[t]{0.45\hsize}
\centering
\includegraphics[width=70mm]{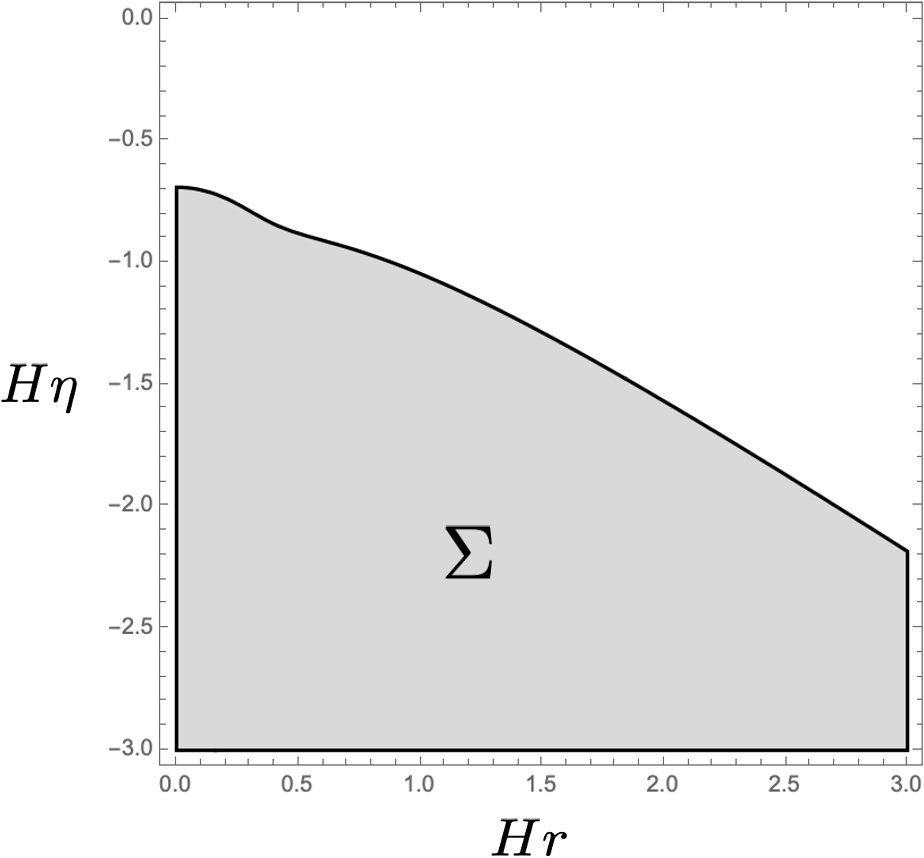}
\caption*{(b)}
\end{minipage} \\\\   

\begin{minipage}[t]{0.45\hsize}
\centering
\includegraphics[width=70mm]{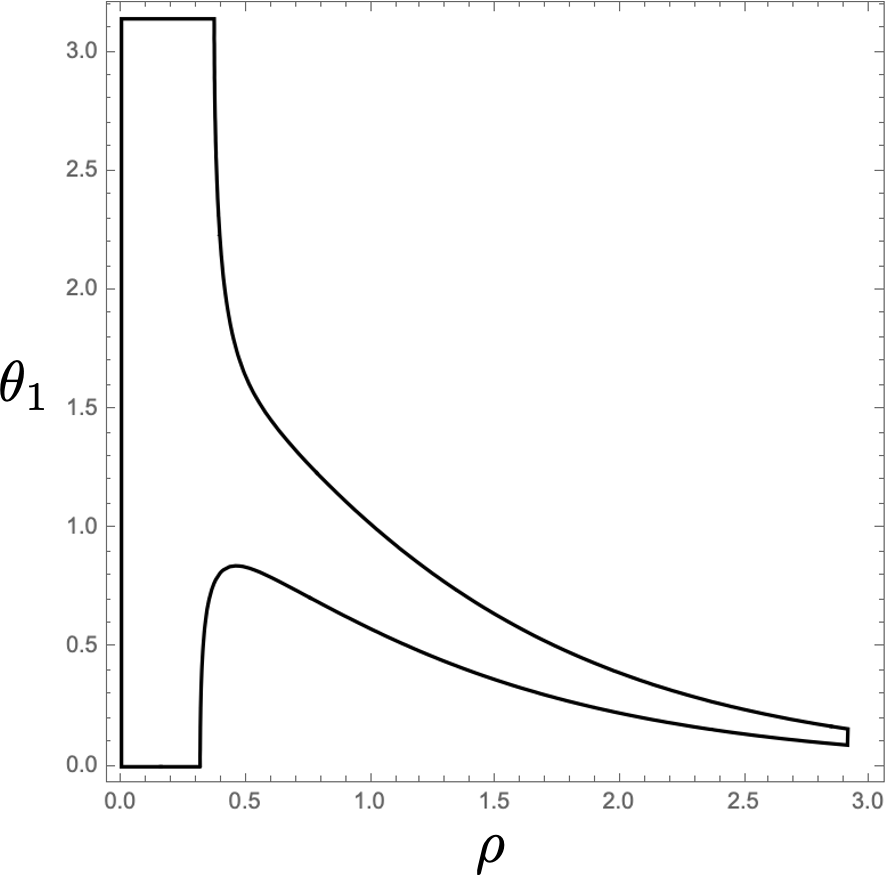}
\caption*{(c)}
\end{minipage} &

\begin{minipage}[t]{0.45\hsize}
\centering
\includegraphics[width=70mm]{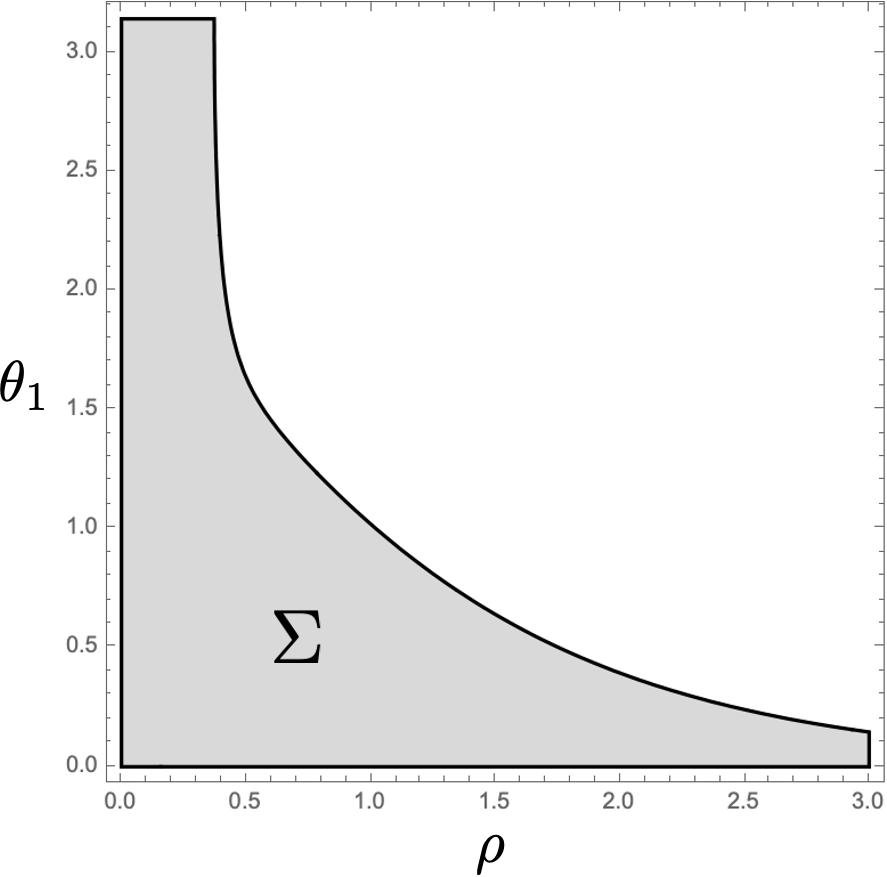}
\caption*{(d)}
\end{minipage} 

\end{tabular}
\caption{(a): The two non-trivial curves on which \eqref{eq:hypersurface} is satisfied. We insert the mesh to confirm whether the curves are spacelike or timelike. It can be seen that the upper line is spacelike but the lower one is partially timelike.
(b): The integration region for \eqref{eq:ActionCDL}. There are another spacelike curve at $\eta=-\infty$.
(c): The two non-trivial curves in $(\rho,\theta_1)$ plane is plotted.
(d): The integration region for \eqref{eq:ActionCDL} in $(\rho,\theta_1)$ plane.
All plots are the case for $\beta=0.5$ and $\alpha g/H=\sqrt{140}.$
}
\label{fig:integral_region}
\end{figure}

We also have to examine the boundary conditions in the Hamilton flow. As in the case of HM tunneling, we expect that there are several intersections where $\phi_{\Delta}=0$ (FIG. \ref{fig:configuration_flow}). Since we have the concrete $\phi_{\Delta}$ \eqref{eq:MomentumZeroCDL} and the bubble configurations \eqref{eq:Fitting}, it is possible to obtain the curves in $(r,\eta)$-plane on which $\phi_{\Delta}=0$. Using \eqref{eq:MomentumZeroCDL} and \eqref{eq:HamiltonEqInhomoZero}, such curves are given as
\begin{equation}
    F(\eta,r):=\partial_{\eta}^{2}\phi_c - \frac{2}{\eta}\partial_{\eta}\phi_c=0.
    \label{eq:hypersurface}
\end{equation}
Note that this is equivalent to solve $a^{-2}\nabla^2\phi_c-V'(\phi_c)=0$.
Substituting \eqref{eq:MomentumZeroCDL} into \eqref{eq:hypersurface} and using the fitting formula \eqref{eq:Fitting}, we find four solutions which are expressed as four lines in the $(\eta,r)$-plane; two of them are placed at $\eta=-\infty$ with fixed $r$ and $r=\infty$ with fixed $\eta$. 
These correspond to $\rho=\infty$ where $\phi=-\alpha$ (FIG. \ref{fig:InhomoBounce}).
Also, we numerically find other two non-trivial hypersurfaces (FIG. \ref{fig:integral_region}(a)). It can be seen that one of the curves is totally spacelike but another is partially timelike. 
Thus, it seems natural to take $\eta=-\infty$ curve as an initial time slice on which $\phi$ is false vacuum and totally spacelike non-trivial curve as a final time slice on which the bubble is nucleated. We denote the region between the two spacelike curves as $\Sigma$.

As a matter of fact, the configurations on the final time slice has a bubble. It can be checked as follows; the value of $\phi_c$ at the origin exceeds the top of potential hill.
In the case of FIG. \ref{fig:integral_region}(a) where we set $(\beta,\alpha g/H)=(0.5,\sqrt{140})$, we have $\phi_c/\alpha\sim -0.1>-0.5=-\beta$ at $Hr=0, H\eta\sim-0.7$.

Now, we can evaluate the action for the bubble configurations for $\Sigma$ (FIG. \ref{fig:integral_region}(b) or (d)).
We use the fitting function \eqref{eq:Fitting} for the numerical integration.
\begin{align}
I
&\simeq \int_{\Sigma} \mathrm{d}^{4}x \left[ - a^3\phi_{\Delta}\left(\ddot{\phi}_c +3H\dot\phi_c\right) - H_\text{CDL}\right]\notag\\
&=-\frac{6\pi}{H^{6}\varepsilon} \int_{\Sigma} \dm\rho \dm\theta_{1} \dm\theta_{2} \dm\theta_{3}
\frac{\sin^{2}\theta_{1} \sin\theta_{2} \sinh^{3}\rho}{(\cosh\rho - \sinh\rho \cos\theta_{1})^{4}} 
\bigg(\partial_{\eta}^{2}\phi_c - \frac{2}{\eta}\partial_{\eta}\phi_c \bigg)^{2} \notag\\
&=-\frac{24\pi^2}{H^{2}\varepsilon} \int_{\Sigma} \dm\rho \dm\theta_{1}
\sin^{2}\theta_{1} \sinh^{3}\rho \bigg[
\frac{(\sinh\rho - \cosh\rho \cos\theta_{1})^{2}}{(\cosh\rho - \sinh\rho \cos\theta_{1})^{2}} \frac{\partial^{2}\phi_c}{\partial \rho^{2}} \notag\\
&\quad\, + \bigg\{
3\frac{\sinh\rho - \cosh\rho \cos\theta_{1}}{\cosh\rho - \sinh\rho \cos\theta_{1}} 
+\frac{\sin^{2}\theta_{1}}{\tanh\rho(\cosh\rho - \sinh\rho \cos\theta_{1})^{2}} \bigg\} 
\frac{\partial \phi_c}{\partial \rho} \bigg]^{2} \notag\\
&:= -\frac{24\pi^2 \alpha^{2}}{H^{2}\varepsilon} \tilde{I}(\mu, \bar{\rho}).
 \label{eq:ActionCDL}
\end{align}
In the second line, we used $H_\text{CDL}=0$, \eqref{eq:MomentumZeroCDL} and \eqref{eq:HamiltonEqInhomoZero}.
Also we performed a coordinate transformation $(t,r)\to(\rho,\theta_1)$. 
In the third line, the following relations are used 
\begin{eqnarray}
-\frac{2}{\eta}\frac{\partial \phi_c(\rho)}{\partial \eta} 
= -\frac{2}{\eta}\frac{\partial \rho}{\partial \eta}\frac{\partial \phi_c(\rho)}{\partial \rho} 
= 2H^{2}(\cosh\rho - \sinh\rho \cos\theta_{1})(\sinh\rho - \cosh\rho \cos\theta_{1})
\frac{\partial\phi_c(\rho)}{\partial \rho},\nonumber
\end{eqnarray}
and 
\begin{eqnarray}
\frac{\partial^{2}\phi_c(\rho)}{\partial\eta^{2}}
&=&H^{2}\bigg[
(\sinh\rho - \cosh\rho \cos\theta_{1})^{2} \frac{\partial^{2}}{\partial \rho^{2}} \notag\\
&&+\bigg\{(\sinh\rho - \cosh\rho \cos\theta_{1})(\cosh\rho - \sinh\rho \cos\theta_{1})
+ \frac{\sin^{2}\theta_{1}}{\tanh\rho}\bigg\} \frac{\partial}{\partial \rho}
\bigg]\phi_c(\rho). \notag
\end{eqnarray}
Here, the partial differentiation with respect to $\eta$ is taken while keeping $r$ constant. 

\subsection{Discussions}
\subsubsection{Appropriate choice of \texorpdfstring{$\varepsilon$}{e}}\label{sec:epsilon}
Since our results \eqref{eq:ActionCDL} depend on $\varepsilon$, we need to choose some specific value of $\varepsilon$ to predict the tunneling rate. We emphasize that the $\varepsilon$ dependence does not imply the pathology of our result. Rather our results should depend on $\varepsilon$ because the non-perturbative effects from UV modes are missed in our formalism as discussed in section~\ref{sec:SK_transition}. 
We then choose $\varepsilon$ as large as possible to evaluate the tunneling rate. 

Our analysis will be valid only when the value of $\varepsilon$ lies in a certain range. Below, we briefly discuss necessary conditions for $\varepsilon$ to justify our analysis. 

\paragraph{Lower bound on $\varepsilon$.}
We need to choose sufficiently large $\varepsilon$ so that IR fields can describe the bubble configuration. This imposes that $(\varepsilon H)^{-1}$ should be smaller than the typical physical size of the bubble or the bubble wall. This condition will be satisfied by taking $\varepsilon\gtrsim \mu$
for our parameter choices. 

\paragraph{Upper bound on $\varepsilon$.}
Our analysis is based on Eq.\eqref{eq:ConditionalFull} which will be invalid when quantum fluctuations of IR fields are too large. 
This imposes an upper bound on $\varepsilon$ because the size of fluctuations of $\phi_c$ at the scale $k\sim k_c(t)$ is proportional to $\varepsilon$:
\begin{equation}
    \delta \phi_c|_{k\sim k_c}
    \coloneqq
    \left[
    \int\frac{\dm^3 k}{(2\pi)^3}\,
    \delta\left(\ln(k/k_c(t))\right)
    \langle 
        |\hat \phi(t,\bm{k})|^2
    \rangle
    \right]^{1/2}
    \approx
    \frac{k_c/a}{2\pi}
    =
    \frac{\varepsilon H}{2\pi}
    \,,
    \label{eq:estimate_qf}
\end{equation}
where we assumed $(k_c/a)^2\gg V''(\phi_c)$~\footnote{This is compatible with the condition $\varepsilon<\alpha/H$ 
when the weak coupling $g\ll1$ is considered.} and approximated $\phi$ as a massless scalar field.

As discussed in section~\ref{sec:SK_transition}, \eqref{eq:ConditionalFull} will be justified when the term $\delta H$ is negligible. 
Then, we define the ratio $\mathcal
Q$ of the term $\delta H$ to the final term on the RHS of \eqref{eq:HamiltonianInhomo} as~\footnote{We can also compare $\delta H$ with the second and the third term on the RHS of \eqref{eq:HamiltonianInhomo}. Such considerations do not change our estimate of $\varepsilon_\text{max}$ in \eqref{eq:boundsummary}.}
\begin{equation}
    \mathcal Q
    \coloneqq
    \left|
    \frac{\delta H}{H^{2}\varepsilon\,a^3\phi_{\Delta}^{2}/(6\pi)}
    \right|_{\phi_\Delta = -\frac{6\pi}{H^{2}\varepsilon}(a^{-2}\nabla^{2}\phi_c - V'(\phi_c))}
    =
    \frac{3\pi^2}{2 \varepsilon^2H^4}
    \left|
        V'''(\phi_c)\left(a^{-2}\nabla^{2}\phi_c - V'(\phi_c)\right)
    \right|
    \,,\no
\end{equation}
and we impose $\mathcal{Q}<1$.
In the second equality, we used $\delta H =(a^3/24)V'''(\phi_c)\phi_\Delta^3$. Similarly to \eqref{eq:estimate_qf}, we can estimate the size of quantum fluctuations of $a^{-2}\nabla^2 \phi$. Then, we estimate $\mathcal Q$ at large $\varepsilon$ as $\mathcal Q\sim g^2\alpha \varepsilon/H$ where we also used $|V'''(\phi_c)|\sim g^2\alpha$ for $|\phi_c| \lesssim \alpha$. 
The condition $\varepsilon<H/(g^2\alpha)$ follows from $\mathcal Q <1$.

\paragraph{Summary.} In summary, it is necessary to choose $\varepsilon$ satisfying the following condition to justify our analysis, 
\begin{equation}
    \mu
    \lesssim
    \varepsilon
    \lesssim\varepsilon_\text{max}
    \coloneqq
    H/(g^2\alpha)
    \approx
    755
    \times
    \left(\frac{1.12\times 10^{-4}}{g}\right)
    \left(\frac{\sqrt{140}}{g\alpha/H}\right)
    \,.\label{eq:boundsummary}
\end{equation}
This suggests that, at least when $g$ is sufficiently tiny, we can reliably choose large $\varepsilon$. Note that we cannot make the action $I$ given in \eqref{eq:ActionCDL} arbitrarily small even if we take $\varepsilon_\text{max}\to\infty$ in the free-theory limit $g\to 0$; for $\varepsilon=\varepsilon_{\text{max}}$, we have $I|_{\varepsilon=\varepsilon_\text{max}}\sim -g^{-1}(g\alpha/H)^3\tilde I(\mu,\bar\rho)$. Roughly speaking, the small $g\alpha/H$ corresponds to the bubble radius being larger than the Hubble radius. Since we are interested in the bubble smaller than the Hubble radius, $g\alpha/H$ cannot be taken small arbitrarily. 

A few more comments on the $\varepsilon$-dependence of our results are in order. 
Our result \eqref{eq:ActionCDL} shows that the value of $I$ becomes larger for smaller choice of $\varepsilon$. This is because the size of quantum fluctuations of modes $k\sim \varepsilon a(t)H$ is proportional to $\varepsilon$, while the nonperturbative physics caused by UV modes $k>\varepsilon a(t)H$ is truncated in our formalism. This interpretation however implicitly assumes that the modes $k\sim \varepsilon_\text{max} a(t) H$ are relevant for the tunneling process. We expect that this is the case because a kinetic energy of fluctuations of modes $k\sim \varepsilon_\text{max} a(t)H$ is estimated as $(\varepsilon_\text{max}H)^4$ which is much smaller than the height of potential barrier $\Delta V \sim g^2\alpha^4$ for our parameter choices.

By contrast, very short-scale physics which is insensitive to the detailed structure of $V(\phi)$ near its origin might be irrelevant for the tunneling process. Hence, we expect that the tunneling rate would converge to some finite value in the limit $\varepsilon\to\infty$. 
It would be interesting to see whether the tunneling rate converges to a finite value as $\varepsilon$ increases, and if so, from which value of $\varepsilon$ this convergence begins.

\subsubsection{Hawking-Moss vs. Coleman-de Luccia}
From the formula \eqref{eq:ConditionalFull}, we derived the two configurations which describe the tunneling from the false vacuum to the true vacuum. One is HM configuration and the other is CDL-like configuration.
Using these configurations, we evaluate the probability of the tunneling.
Here, we compare these probabilities.
Denoting the action \eqref{eq:ActionCDL} and \eqref{eq:ActionHMSY} as $I_{\text{bubble}}$ and $I_{\text{HM}}$, the ratio of the two actions is given by
\begin{align}
\gamma:= \frac{I_{\text{bubble}}}{I_{\text{HM}}} 
= \frac{9\alpha^{2}H^{2}}{ \Delta V \varepsilon} \tilde{I}(\mu, \bar{\rho})
= \frac{1}{\varepsilon} \frac{108}{(1-\beta)^{3}(\beta+3)}\frac{H^{2}}{\alpha^{2}g^{2}} \tilde{I}(\mu, \bar{\rho}) \label{eq:ratio}
\end{align}
For example, if we choose $\beta=0.5$ and $\alpha g/H=\sqrt{140}$ (see FIG. \ref{fig:InhomoBounce}), the fitting parameters become $\mu=7.32515$ and $\bar{\rho}=0.34397$ and the ratio becomes $\gamma = 12.9840/\varepsilon$. 
For these parameters, the upper bound of $\epsilon$ is given as \eqref{eq:boundsummary} and then $\gamma$ can be smaller than one for appropriately small g.
This means that the CDL bubble configuration is dominant rather than HM configuration.
Results for other parameters are also shown in Table \ref{table:BenchMark}.

\subsubsection{Comparison with the Euclidean method}

We also compute the CDL action in Euclidean method (Appendix \ref{sec:Euclidean_CDL_appendix}) and compare the ratios of the CDL action to the HM action (Table \ref{table:BenchMark}). 
For $\varepsilon>\varepsilon_*$, our tunneling process becomes more probable than the one predicted by the Euclidean method. As discussed in section~\ref{sec:epsilon}, we can consistently choose $\varepsilon>\varepsilon_*$ for certain choices of parameters $(g,\alpha, \beta)$. Perhaps, our results may indicate the presence of tunneling process which is more probable than the Euclidean method. 

\begin{table}[bt]
\centering
\begin{tabular}{|c|c||c||c|c||c||c||c|} \hline
$\beta$ & $g^{2}\alpha^{2}/H^{2}$ & $1-\phi_c(10^{-15})/\alpha$ & $\mu$ & $\bar{\rho}$ & $\varepsilon\gamma$ & $\gamma^E$ & $\epsilon_*$\\ \hline\hline
0.3 & 80 & 4.87210$\times$ $10^{-5}$ & 6.22563 & 1.05197 & 355.089 & 0.247004 & 1437.58 \\ \hline
0.4 & 80 & 5.20273$\times$ $10^{-3}$ & 5.99645 & 0.65257 & 64.1185 & 0.0909796 & 704.758 \\ \hline
0.5 & 80 & 3.89106$\times$ $10^{-2}$ & 5.63444 & 0.475247 & 33.9526 & 0.0599470 & 566.376 \\ 
\hline\hline
0.3 & 140 & 3.50945$\times$ $10^{-4}$ & 8.13990 & 0.673305 & 52.2515 & 0.0486678 & 1073.64 \\ \hline
0.4 & 140 & 8.96003$\times$ $10^{-3}$ & 7.82419 & 0.459659 & 19.2912 & 0.0239903 & 804.123\\ \hline
0.5 & 140 & 4.98284$\times$ $10^{-2}$ & 7.32515 & 0.34397 & 12.9840 & 0.0172063 & 754.606 \\ \hline
\end{tabular}
\caption{The values of $\gamma$ for several parameter sets. First and second ones are potential parameters. Third one is the initial condition for the equation of motion \eqref{eq:HamiltonEqO4}. Forth and fifth ones are fitting parameters for \eqref{eq:Fitting}. Sixth and seventh ones are the numerical results of \eqref{eq:ratio} and \eqref{eq:ratio_E}. The last one is the cut-off parameter $\epsilon$ for which \eqref{eq:ratio} coincides with \eqref{eq:ratio_E}.}
\label{table:BenchMark}
\end{table}

\subsubsection{Bubble nucleation hypersurface and the subsequent evolution}

We find the non-trivial spacelike hypersurface on which the bubble is nucleated (FIG. \ref{fig:integral_region}(a)), where the hypersurface is given by the condition $F(\eta,r)=0$ or equivalently $a^{-2}\nabla^2\phi_c-V'(\phi_c)=0$.
The subsequent evolution is then described by the one shown in the first line of \eqref{eq:HamiltonEqInhomoZero} which is the standard classical equation of motion. The configuration on the hypersurface gives the initial data for the classical dynamics after the bubble nucleation. 

The field value on the spacelike hypersurface does not reach the true vacuum but rather it lies between the true vacuum and the top of the potential as mentioned in the subsection \ref{sec:bubble_sol}. This is possible because the location of hypersurface is defined by the condition  $a^{-2}\nabla^2\phi_c-V'(\phi_c)=0$ where the gradient force is balanced with the potential force. Though it is non-trivial to solve the evolution starting from the general spacelike hypersurfaces, it is desirable to solve it to
fully understand the formation of the true-vacuum bubble in our scenario. 

We conclude this section by pointing out that our formalism naturally predicts that the condition $\overline\phi_\Delta=0$ (or $\phi_\Delta=0$ via \eqref{eq:noise_CDL}) defines the hypersurface on which the quantum dynamics is switched to the classical dynamics and the non-trivial field configuration is nucleated. This would be the generic prediction of our formalism that holds true once the appropriate configuration $\overline{\phi}_\Delta$ is specified.

\section{Conclusion}\label{sec:concl}
We studied the tunneling processes on de Sitter background by using the stochastic approach. 
A novel point is that we applied the MSRJD functional integral to the problem. In this formalism, the tunneling rate is obtained by evaluating the functional integral with the saddle-point approximation. Using this method, we investigated both the HM and the CDL tunnelings.

For the HM case, we first analyzed \eqref{eq:ConditionalSY} which is the MSRJD functional integral of the stochastic equation \eqref{eq:LangevinSY} on a single spatial point. We succeeded in deriving the tunneling solution in the ``phase space'' (FIG. \ref{fig:HamiltonFlow}) which represents the tunneling process from the false vacuum, through the top of the potential hill, and finally to the true vacuum (FIG. \ref{fig:HomoBounce}). This solution has natural tunneling boundary conditions which cannot be obtained from the Euclidean method. 
The tunneling rate for this configuration coincides with that of the HM instanton at the leading order. In sec.~\ref{subsec:HM_full}, we succeeded in re-deriving this result starting from \eqref{eq:ConditionalFull} which describes the stochastic dynamics in the global region covered by the flat chart. We also estimated the time scale of the HM tunneling as \eqref{timescale} which becomes much longer than the Hubble time scale. Our analysis clarifies the physical picture of the HM instanton, i.e., the HM transition probability represents the transition probability of a coarse-grained patch with a physical radius $(\varepsilon H)^{-1}$. Some technical complications that arise in dealing with \eqref{eq:ConditionalFull} and the way of handling them are remarked in sec.~\ref{sec:remark}. Physically, these complications are due to the spatial correlations of stochastic noises. 

For the CDL case, we found the configurations which describe the bubble nucleation process.
The tunneling rates for the bubble configurations depend on the cutoff scale dividing IR and UV fields. We argued that this dependence comes from the truncation of the non-perturbative effects from UV modes. We also discussed the valid choice of the cutoff scale for which the MSRJD functional integral will be reliable. This consideration is based on the first-principles derivation of stochastic approach from the Schwinger-Keldysh formalism. With an appropriate cutoff scale, it turned out that the CDL tunneling rate is larger than the HM one for the steep potential barrier. However, as mentioned above \eqref{eq:HamiltonianInhomo}, it has not yet been investigated if our configuration can be really used for estimating the tunneling action. This aspect would be important for evaluating the tunneling action, being left for future work. Nonetheless, we believe that our study clarifies how we can proceed the analysis and how the bubble nucleation process could be described in the stochastic approach; for instance, our formalism can naturally define the location of hypersurface on which the quantum dynamics is switched to the classical dynamics and the non-trivial field configuration is nucleated.

It is interesting to understand the relation between the stochastic approach and the Euclidean method. 
We compared the CDL tunneling rate in our method with that in Euclidean method and found the former becomes larger than the latter for the potential considered with a certain cutoff scale. It would be worth investigating the meaning of this result. 
Especially, we need to know a precise relation between the CDL instanton and our configuration.
Intriguingly, our method is also related to that discussed in \cite{Braden:2018tky,Blanco-Pillado:2019xny,Hertzberg:2020tqa,Tranberg:2022noe,Hertzberg:2019wgx}, where the bubble accidentally appears from the initial quantum fluctuations.
In our method, however, we assume that we can neglect the non-perturbative effects from UV modes, the quantum component of the potential of the form $V'''(\phi_c)\phi_\Delta^3$, and that the configuration \eqref{eq:MomentumZeroCDL} can be used to estimate the tunneling rate. The first two are summarized into the term $\delta H$. We also assume the fixed background spacetime. The relaxation of these assumptions would be important to seek the relations among the three methods. 
It would also be useful to apply our path-integral method to the tunneling in flat space and make a comparison with the Euclidean method.

Once this relaxation is achieved, it would also be interesting to apply our method to tunneling phenomena where the backreaction to the background geometry is non-negligible and make comparisons with recently-discussed methods. For instance, the formalism of the Wheeler-DeWitt equation \cite{Kristiano:2018oyv, Cespedes:2020xpn,Maniccia:2022iqa} and the tunneling potential \cite{Espinosa:2018voj} must be related to ours because the HM exponent is reproduced. The tunneling in the black hole spacetime \cite{Gregory:2020cvy,Gregory:2020hia} is also interesting. Especially, \cite{Gregory:2020hia} discussed the HM transition with a black hole from the viewpoint of the stochastic approach.

Our method using the saddle-point approximation effectively amounts to solving real-time quantum dynamics, which reduces to Starobinsky's stochastic approach~\cite{Starobinsky:1986fx,Starobinsky:1994bd} at leading order when the super-horizon dynamics in de Sitter space is considered. However, since our starting point is the Schwinger-Keldysh path integral, which can be formulated in a more generic setup and incorporates all the quantum effects in principle, we believe that our formalism sheds light on further studies of tunneling phenomena from a real-time perspective.

One of the advantage of our method is that it is applicable to a dynamical setup.
The application of our method to the inflation models such as the chain inflation \cite{Freese:2004vs} and the warm inflation \cite{Berera:1995ie} is also intriguing.
 Using our method, we can study the tunneling process in the inflationary background under dissipation and fluctuations coming from circumference. 
 We leave these issues for future work.

\section*{Author Contributions:}
 Conceptualization, T.M., J.S. and J.T.; Methodology, T.M., J.S. and J.T.; Software, T.M.; Investigation, T.M., J.S. and J.T.; Writing – original draft, T.M., J.S. and J.T.; Visualization, T.M.
All authors have read and agreed to the published version of the manuscript.

\section*{Funding:}
T. M. was supported by JST SPRING, Grant Number JPMJSP2148 and JSPS KAKENHI Grant Number JP23KJ1543.
J.\ S. was in part supported by JSPS KAKENHI Grant Numbers JP17H02894, JP17K18778, JP20H01902, JP22H01220.
J.T. is supported by IBS under the project code, IBS-R018-D1.

\section*{Data Availability Statement:}
Data are contained within the article.

\section*{Conflicts of Interest:}
The authors declare no conflict of interest.

\appendix
\section{Stochastic approach from the first principle}\label{sec:in-in}
In this section, we evaluate the generating functional for IR fields $Z[J^\text{IR}(T)]$ defined in Eq.\eqref{IRgene_def} to obtain \eqref{tunnel_IRgene_main}. Note that each step we need to take for evaluating $Z[J^\text{IR}(T)]$ is summarized in section~\ref{sec:SK_transition}.

\subsection{Path integral representation}
A path integral representation of $Z[J^\text{IR}(T)]$ takes the following standard form,
\begin{align}
    Z[J^\text{IR}(T)]
    = &
   \int\mathcal{D}(\phi_+,\phi_-,\Pi_+,\Pi_-)
    \, e^{i \bm{J}^\text{IR}\cdot(\bm\phi^\text{IR}_++\bm\phi^\text{IR}_-)/2}
    \no\\
    &\quad\times
    e^{i(S^+_H-S^-_H)}
    \prod_{\bm x}\delta(\phi_+(T,\bm x)-\phi_-(T,\bm x))
    \Psi_0[\phi_+]\Psi^*_0[\phi_-]
    \,,\label{IRgene1}
\end{align}
where $\Psi_0[\phi]$ is an initial wave functional. We specify it more concretely in section~\ref{subsubsec:UVintegral}.
$S_H^\pm=\int^{t=T}\mathrm{d}^4x (\Pi^\pm\dot{\phi}^\pm-H[\phi^\pm,\Pi^\pm])$ is a Hamiltonian action. After the rotation of the basis $(X_+,X_-)\to(X_c,X_\Delta)\coloneqq(\frac{X_++X_-}{2},X_+-X_-)$ with $X=\phi,\Pi$, \eqref{IRgene1} is written as
\begin{align}
    Z[J^\text{IR}(T)]
    = &
   \int\mathcal{D}(\phi_c,\phi_\Delta,\Pi_c,\Pi_\Delta)
    \, e^{i \bm{J}^\text{IR}\cdot\bm\phi^\text{IR}_c}
    e^{iS_H[\phi_c,\phi_\Delta,\Pi_c,\Pi_\Delta]}
    \prod_{\bm x}\delta(\phi_\Delta(T,\bm x))
    \rho_0[\phi_c,\phi_\Delta]
    \,.\label{IRgene2}
\end{align}
Here, we defined $S_H[\phi_c,\phi_\Delta,\Pi_c,\Pi_\Delta]\coloneqq (S_H^+-S_H^-)|_{(X_+,X_-)\to(X_c,X_\Delta)}$ with $X=\phi,\Pi$. We also defined $\rho_0[\phi_c,\phi_\Delta]
\coloneqq \Psi_0[\phi_c+(\phi_\Delta/2)]\Psi^*_0[\phi_c-(\phi_\Delta/2)]$.

\subsection{Nonperturbative generating functional for IR sector}
Now we explain how to calculate $Z[J^\text{IR}(T)]$ non-perturbatively in the IR sector. 
This allows us to capture the non-perturbative physics of the sector $k\leq k_c(t)$. Our strategy is to split the integration variables into UV modes $k> k_c(t)$ and IR modes $k\leq k_c(t)$ for each time step $t$, and perform the integration over UV variables to get $Z[J^\text{IR}(T)]$.
One may simply split the integration variables $X_a(t,\bm k)$,\footnote{In the main text, we use the notation $X_{\bm k}(t)$ to write variables in momentum space. We however adopt the notation $X(t,\bm k)$ in this appendix since we have many subscripts such as $c$ and $\Delta$.} 
where $X=(\phi,\Pi)$ and $a=(c,\Delta)$, by the following replacement in \eqref{IRgene2}:
\begin{equation}
X_a(t,\bm k) \to
\begin{cases}
  X_a^\text{IR}(t,\bm k)  & (t\geq t_k) \\
  X_a^\text{UV}(t,\bm k)  & (t< t_k)\,,
\end{cases}
\label{naive_repl}
\end{equation}
where $t_k$ is defined by the condition $k_c(t_k)=k$.
It is then tempted to perform the integration over UV variables $X^\text{UV}_a$ by using the Schwinger-Keldysh (or {\it  closed-time-path}) formalism.
However, the expression obtained from \eqref{IRgene2} after the replacement \eqref{naive_repl} does not have the product of delta functions $\prod_{\bm k}\delta(\phi^\text{UV}_\Delta(t_k-0^+,\bm k))$ at the final time $t=t_k-0^+$.\footnote{Here, $0^+$ is the infinitesimal time step which is introduced to obtain path integral representation of unitary time evolution as usual.}
That is, the time contour for each UV $\phi$-variable with modes $0<k\leq k_c(T)$ is not closed at the final time. Hence, the usual Schwinger-Keldysh formalism does not apply for evaluating the integration over UV variables.
To resolve this issue, Refs.\cite{Tokuda:2017fdh,Tokuda:2018eqs} proposed a new way of splitting integration variables. 
 The splitting is given by \eqref{naive_repl} for $X=\Pi$, while for $\phi$-variable it is defined by the following rule rather than \eqref{naive_repl}:
\begin{equation}
\begin{array}{lr}
\phi_c(t,\bm k)\to
\begin{cases}
\phi_{c}^\text{IR}(t,\bm k)& (t>t_k)\,\\
\phi_{c}^\text{UV}(t,\bm k)& (t\leq t_k)\,,
\end{cases} 
& \quad
\phi_{\Delta}(t,\bm k)\rightarrow
\begin{cases}
\phi_{\Delta}^\text{IR}(t,\bm k)& \quad (t\geq t_k) \,\\
\phi_{\Delta}^\text{UV}(t,\bm k)& \quad (t<t_k)\,.\label{repl_phi}
\end{cases}
\end{array}
\end{equation}
In this replacement~\eqref{repl_phi}, the variables $\phi^\text{UV}_\Delta(t_k,\bm k)$ are absent while the variables $\phi^\text{UV}_c(t_k,\bm k)$ are present. Therefore, in the expression obtained from \eqref{IRgene2} after the new replacement, the time contour for every UV $\phi$-variable including those for modes $\bm k$ with $0<k\leq k_c(T)$ is closed. This allows us to perform the integration over UV variables based on the Schwinger-Keldysh formalism as we see below. Note that there is no subtlety in the replacement of integration variables $X_a(t,\bm k)$ for a zero mode $k=0$ and those for deep UV modes $k>k_c(T)$: in both replacement rules mentioned above, they are simply replaced by IR variables $X_a^\text{IR}(t,\bm k)$ and UV variables $X_a^\text{UV}(t,\bm k)$, respectively. The time contours for them are closed thanks to the product of delta functions in the original expression \eqref{IRgene2}.

After the new replacement, the term $S_H[\phi_c,\phi_\Delta,\Pi_c,\Pi_\Delta]$ in the exponent of \eqref{IRgene2} is decomposed into three pieces: purely IR terms $S_H^\text{IR}\coloneqq S_H[\phi^\text{IR}_c,\phi^\text{IR}_\Delta,\Pi^\text{IR}_c,\Pi^\text{IR}_\Delta]$, purely UV terms $S_H^\text{UV}\coloneqq S_H[\phi^\text{UV}_c,\phi^\text{UV}_\Delta,\Pi^\text{UV}_c,\Pi^\text{UV}_\Delta]$, and the remaining IR-UV mixing terms $S_\text{mixed}$:
\begin{equation}
    S_H[\phi_c,\phi_\Delta,\Pi_c,\Pi_\Delta]
    \to
    S_H^\text{IR} + S_H^\text{UV} + S_\text{mixed}\,.
\end{equation}
In terms of these quantities, $Z[J^\text{IR}(T)]$ can be formally written as~\cite{Tokuda:2017fdh,Tokuda:2018eqs} 
\begin{align}
    Z[J^\text{IR}(T)]
    =
    &\int\mathcal{D}(\phi^\text{IR}_c,\phi^\text{IR}_\Delta,\Pi^\text{IR}_c,\Pi^\text{IR}_\Delta)
    \,
    e^{i \bm{J}^\text{IR}\cdot\bm\phi^\text{IR}_c}
    e^{iS_H^\text{IR}}e^{i\Gamma}\notag\\
    &\times
    \prod_{k\leq k_c(T)}
    \delta(\phi^\text{IR}_\Delta(T,\bm k))
    \prod_{0<k\leq k_c(T)}
    \delta(\phi^\text{IR}_c(t_k,\bm k)),
    \label{IRgene3}
\end{align}
where $i\Gamma$ is the effective action for IR fields,
\begin{equation}
    e^{i\Gamma}
    \coloneqq
    \int\mathcal{D}
    (\phi^\text{UV}_c,\phi^\text{UV}_\Delta,\Pi^\text{UV}_c,\Pi^\text{UV}_\Delta)
    \,
    e^{i(S_H^\text{UV}+S_\text{mixed})}
   \prod_{k>0}\delta(\phi^\text{IR}_\Delta(t_f(k),\bm k)) \rho_0
   [\phi^\text{UV}_c,\phi^\text{UV}_\Delta;\phi^\text{IR}_c,\phi^\text{IR}_\Delta]
    \,,\label{infl1}
\end{equation}
with $t_f(k)\coloneqq \min[T,t_k]$. Here, $\rho_0[\phi_c,\phi_\Delta]\to \rho_0
[\phi^\text{UV}_c,\phi^\text{UV}_\Delta;\phi^\text{IR}_c,\phi^\text{IR}_\Delta]$ due to the replacement of variables. The IR variables $(\phi^\text{IR}_c,\phi^\text{IR}_\Delta)$ in the argument of $\rho_0$ contain only a zero mode.
Since the time path for each UV $\phi$-variable is closed in \eqref{infl1}, we can calculate $i\Gamma$ perturbatively as usual by writing the connected diagrams with $n$ external IR fields that are connected by UV propagators. The diagramatic rules such as vertex factors and symmetry factors follow the standard rules of Scheinger-Keldysh formalism for given vertexes in $S_H^\text{UV}+S_\text{mixed}$.

Interestingly, due to the non-trivial replacement rule \eqref{repl_phi}, the $\Pi\dot\phi$ terms in $S_H$ also contribute to $S_\text{mixed}$~\cite{Morikawa:1989xz,Tolley:2008qv,Tokuda:2017fdh,Tokuda:2018eqs}. Referring to such contributions and the remaining terms in $S_\text{mixed}$ as $S_\text{tr}$ and $S_\text{mix-int}$, respectively, we have $S_\text{mixed}=S_\text{tr}+S_\text{mix-int}$ with 
\footnote{Precisely speaking, we need to perform the UV-IR splitting in the path integral with infinitesimal discrete time step for deriving $S_\text{tr}$ correctly. After taking the continuum limit, we obtain \eqref{bilinear}; see also \cite{Tokuda:2018eqs}. Furthermore, the form of $S_\text{mix-int}$ depends on the model. In our case, we have 
\begin{align}
    S_\text{mix-int}
    =
    \int\mathrm{d}^4x
    &\bigl[
        V(\phi^\text{UV}_++\phi^\text{IR}_+)-V(\phi^\text{UV}_+) - V(\phi^\text{IR}_+)
        \no\\
        &
        - V(\phi^\text{UV}_-+\phi^\text{IR}_-)+V(\phi^\text{UV}_-) + V(\phi^\text{IR}_-)
    \bigr]_{\phi^\text{UV}_\pm=\phi^\text{UV}_c \pm (\phi^\text{UV}_\Delta/2),\,\phi^\text{IR}_\pm=\phi^\text{IR}_c \pm (\phi^\text{IR}_\Delta/2)}
    \,.\label{nonlinear}
\end{align}}
\begin{align}
    S_\text{tr}
    \coloneqq
    \int^T_{-\infty}\mathrm{d}t\int\frac{\mathrm{d}^3k}{(2\pi)^3}
    \delta(t-t_k)
    \left[
        \phi^\text{IR}_\Delta(t,-\bm k)\Pi^\text{UV}_c(t,\bm k)
        -
        \Pi^\text{IR}_\Delta(-\bm k,t)\phi^\text{UV}_c(t,\bm k)
    \right]
    \,.
    \label{bilinear}
\end{align}
The bi-linear vertexes exist only at the UV$\to$IR transition time $t_k$ for given modes with $0< k \leq k_c(T)$. Physically, these vertexes set the initial fluctuations for such modes in \eqref{IRgene3}.  

\subsubsection{Integrate out short-wavelength modes}\label{subsubsec:UVintegral}
Now we calculate $i\Gamma$ perturbatively by specifying the setup more concretely.  
We organize the perturbation theory around the false vacuum state $\rho_0$ for which the expectation value of the zero mode is $\phi_\text{false}$.  
We assume that the fluctuations of zero modes are tiny enough so that the couplings between zero-mode fluctuations and those of non-zero modes $k>0$ are switched off in the past infinity $t=-\infty$. We then take the initial state of non-zero modes to be the Bunch-Davies vacuum state for a free field, where the time evolution of a free field is defined by the quadratic action expanded around the homogeneous background $\phi_\text{false}$. 
These assumptions can be implemented by writing the initial quantum state as
\begin{equation}
    \rho_0[\phi^\text{UV}_c,\phi^\text{UV}_\Delta;\phi^\text{IR}_c,\phi^\text{IR}_\Delta]
    =
    \rho_\text{BD}[\phi^\text{UV}_c,\phi^\text{UV}_\Delta;\phi_\text{false}]\,
    \rho_0[\phi^\text{IR}_c-\phi_\text{false},\phi^\text{IR}_\Delta]
    \label{initial_state}
\end{equation}
which is properly normalized as 
\begin{align}
    1
    &=\prod_{k>0}\int\mathrm{d}\phi^\text{UV}_c(-\infty,\bm k)\int\mathrm{d}\phi^\text{UV}_\Delta (-\infty,\bm k)\,
    \rho_\text{BD}[\phi^\text{UV}_c,\phi^\text{UV}_\Delta;\phi_\text{false}]
    \no\\
    &=
    \int\mathrm{d}\phi^\text{IR}_c(-\infty,\bm 0)\int\mathrm{d}\phi^\text{IR}_\Delta(-\infty,\bm0)\,
    \rho_0[\phi^\text{IR}_c-\phi_\text{false},\phi^\text{IR}_\Delta]\,.
\end{align}
$\rho_\text{BD}[\phi^\text{UV}_c,\phi^\text{UV}_\Delta;\phi_\text{false}]$ is the density matrix for non-zero modes. An initial state for the zero mode is given by $ \rho_0[\phi^\text{IR}_c-\phi_\text{false},\phi^\text{IR}_\Delta]$ for which we have
\begin{equation}
    \bra{\Psi(-\infty)}\hat\phi(\bm k = \bm 0)\ket{\Psi(-\infty)}
    =
    \int\mathrm{d}\phi^\text{IR}_c 
    \rho_0[\phi^\text{IR}_c-\phi_\text{false},\phi^\text{IR}_\Delta] 
    \phi^\text{IR}_c
    =
    \phi_\text{false}(2\pi)^3\delta(\bm 0)
    \,.\label{vev_zeromode1}
\end{equation}
Combining this with $\bra{\Psi(-\infty)}\hat\phi(\bm k)\ket{\Psi(-\infty)}=0$ for non-zero modes $k>0$, we have 
\begin{equation}
    \bra{\Psi(-\infty)}\hat\phi(\bm k)\ket{\Psi(-\infty)}
    =
    \phi_\text{false}(2\pi)^3\delta(\bm k)
    \,.\label{vev_momspace}
\end{equation}
Hence, the VEV of the IR field $\hat\phi^\text{IR}(t,\bm x)$ in the past infinity $t=-\infty$ is given by $\phi_\text{false}$.

The interaction vertexes for calculating $i\Gamma$ are the bi-linear vertexes \eqref{bilinear} and non-linear vertexes coming from interacting potential $V(\phi)$. We assume that the potential is sufficiently flat under the region we are interested in. It is then natural to separate $i\Gamma$ into the leading order pieces and higher-order terms in the coupling constant as follows,
\begin{align}
&   e^{i\Gamma}
    = 
    \rho_0[\phi^\text{IR}_c-\phi_\text{false},\phi^\text{IR}_\Delta] \,e^{i\Gamma_\text{LO}^{k>0}}
    e^{-i\int\mathrm{d}^4x\,\delta H_\text{higher}}
    \,,
    \no\\
    &
    e^{i\Gamma^{k>0}_\text{LO}}
    \coloneqq
   \int\mathcal{D}
    (\phi^\text{UV}_c,\phi^\text{UV}_\Delta,\Pi^\text{UV}_c,\Pi^\text{UV}_\Delta)
    \,
    e^{iS^\text{UV}_{H,\text{free}}
    +S_\text{tr}}
    \prod_{k>0}\delta(\phi^\text{IR}_\Delta(t_f(k),\bm k))
    \rho_\text{BD}[\phi^\text{UV}_c,\phi^\text{UV}_\Delta;\phi_\text{false}]
    \,.\label{infl2}
\end{align}
Here, $S^\text{UV}_{H,\text{free}}[\phi^\text{UV}_c,\phi^\text{UV}_\Delta,\Pi^\text{UV}_c,\Pi^\text{UV}_\Delta;\phi_\text{false}]$ is the free part of the Hamiltonian action defined around the false vacuum as explained above. Higher-order corrections to $\Gamma$ in the coupling constant are represented by the term $\delta H_\text{higher}$ which can be evaluated perturbatively. We have $\delta H_\text{higher}=0$ at the leading order. 

To calculate $i\Gamma^{k>0}_\text{LO}$ by writing down the connected diagrams with external IR fields, we only need to consider the bi-linear vertexes $S_\text{tr}$ given in \eqref{bilinear}. 
The number of diagrams are only four, and $i\Gamma_\text{LO}^{k>0}$ can be calculated as~\cite{Tokuda:2017fdh,Tokuda:2018eqs}
\begin{align}
    i\Gamma_\text{LO}^{k>0}
    &=
    \frac{-1}{2}\int\mathrm{d}t\int\mathrm{d}t'\int\frac{\mathrm{d}^3k}{(2\pi)^3}\,
    X_\alpha(t,\bm k) g^{\alpha\beta}(t,t',k)X_\beta(t',-\bm k)
    \no\\
    &=
    \frac{-1}{2}\int\mathrm{d}^4x\int\mathrm{d}^4x'\,
    X_\alpha(x) G^{\alpha\beta}(x,x')X_\beta(x')
    \,.\label{infl3}
\end{align}
Here, the Greek indices $\alpha,\beta=(\phi,\Pi)$ label the IR-$\Delta$ fields: $(X_\phi,X_\Pi)=(\Pi^\text{IR}_\Delta, -\phi^\text{IR}_\Delta)$.
Substituting Eq.\eqref{infl2} into \eqref{IRgene3}, we find
\begin{align}
    Z[J^\text{IR}(T)]
    =
    &\int\mathcal{D}(\phi^\text{IR}_c,\phi^\text{IR}_\Delta,\Pi^\text{IR}_c,\Pi^\text{IR}_\Delta)
    \,
    e^{i \bm{J}^\text{IR}\cdot\bm\phi^\text{IR}_c}
    \rho_0[\phi^\text{IR}_c-\phi_\text{false},\phi^\text{IR}_\Delta]\,e^{i(S_H^\text{IR}+\Gamma_\text{LO}^{k>0})} e^{-i\int\mathrm{d}^4x\,\delta H_\text{higher}}
    \no\\
    & \times
    \prod_{k\leq k_c(T)}
    \delta(\phi^\text{IR}_\Delta(T,\bm k))
    \prod_{0<k\leq k_c(T)}
    \delta(\phi^\text{IR}_c(t_k,\bm k))
   \,.
    \label{IRgene4}
\end{align}
Now we introduce $i\Gamma_{k=0}$ as 
\begin{align}
    e^{i\Gamma_{k=0}[\phi^\text{IR}_\Delta,\Pi^\text{IR}_\Delta]}
    \coloneqq
    \int\mathrm{d}\phi^\text{IR}_c
    \rho_0[\phi^\text{IR}_c-\phi_\text{false},\phi^\text{IR}_\Delta]e^{-i\Pi^\text{IR}_\Delta(\phi^\text{IR}_c-\phi_\text{false}(2\pi)^3\delta(\bm0))}
    \,.
\end{align}
By definition, we have
\begin{align}
    &\int\mathrm{d}\phi^\text{IR}_c
    \rho_0[\phi^\text{IR}_c-\phi_\text{false},\phi^\text{IR}_\Delta]
    \,
    e^{
    -i\Pi^\text{IR}_\Delta
    \left(
        \phi^\text{IR}_c-\phi_\text{false}(2\pi)^3\delta(\bm0)
    \right)
    }
    \no\\
    &=
    \int\mathrm{d}\phi^\text{IR}_c\,
    e^{-i\Pi^\text{IR}_\Delta\phi^\text{IR}_c} e^{i\Gamma_{k=0}[\phi^\text{IR}_\Delta,\Pi^\text{IR}_\Delta]}
    \delta
    \left(
        \phi^\text{IR}_c-\phi_\text{false}(2\pi)^3\delta(\bm0)
    \right)
    \,.\label{trick1}
\end{align}
Substituting \eqref{trick1} into \eqref{IRgene4}, we find
\begin{align}
    Z[J^\text{IR}(T)]
    =
    &\int\mathcal{D}(\phi^\text{IR}_c,\phi^\text{IR}_\Delta,\Pi^\text{IR}_c,\Pi^\text{IR}_\Delta)
    \,
    e^{i \bm{J}^\text{IR}\cdot\bm\phi^\text{IR}_c}
    \,e^{i(S_H^\text{IR}+\Gamma_\text{LO}^{k>0}+\Gamma_{k=0})}
     e^{-i\int\mathrm{d}^4x\,\delta H_\text{higher}}
     \no\\
    &\times
    \prod_{k\leq k_c(T)}
    \delta(\phi^\text{IR}_\Delta(T,\bm k))
    \delta(\phi^\text{IR}_c(t_k,\bm k)-\phi_\text{false}(2\pi)^3\delta(\bm k))
   \,,
    \label{IRgene5}
\end{align}
with defining $t_k|_{k=0}=-\infty$. 

\subsubsection{Long-wavelength sector}
For later convenience, let us focus on the term $\int\mathrm{d}t\,\Pi^\text{IR}_c(t,\bm k)\dot\phi^\text{IR}_\Delta(t,-\bm k)$ in $S_H^\text{IR}$. We use the following trick to perform the integration by parts:
\begin{align}
    &\delta(\phi^\text{IR}_\Delta(T,\bm k))
    \,e^{i
    \int^T_{t_k}\mathrm{d}t\,\Pi^\text{IR}_c(t,\bm k)\dot\phi^\text{IR}_\Delta(t,-\bm k)}
    \no\\
    &= 
    \delta(\phi^\text{IR}_\Delta(T,\bm k))
    \int \mathrm{d}\Pi^\text{IR}_c(t_k-0^+,\bm k)
    \,e^{i
    \int^T_{t_k-0^+}\mathrm{d}t\,
    \Pi^\text{IR}_c(t,\bm k)\dot\phi^\text{IR}_\Delta(t,-\bm k)} 
    \delta(\Pi^\text{IR}_c(t_k-0^+,\bm k))
    \no\\
    &=
    \delta(\phi^\text{IR}_\Delta(T,\bm k))
    \int \mathrm{d}\Pi^\text{IR}_c(t_k-0^+,\bm k) 
    \,e^{-i
    \int^T_{t_k-0^+}\mathrm{d}t\,
    \dot\Pi^\text{IR}_c(t,\bm k)\phi^\text{IR}_\Delta(t,-\bm k)} 
    \delta(\Pi^\text{IR}_c(t_k-0^+,\bm k))
    \,.\label{trick2}
\end{align}
Thanks to this trick, \eqref{IRgene5} can be written as 
\begin{align}
    Z[J^\text{IR}(T)]
    &=
    \int\mathcal{D}(\phi^\text{IR}_c,\phi^\text{IR}_\Delta,\Pi^\text{IR}_c,\Pi^\text{IR}_\Delta)
    \int\widetilde{\mathcal{D}}\Pi^\text{IR}_c
    \,
    e^{i \bm{J}^\text{IR}\cdot\bm\phi^\text{IR}_c}
    \,e^{i(S_{H,(d)}^\text{IR}+S_{H,(s)}^\text{IR}+\Gamma_\text{LO}^{k>0}+\Gamma_{k=0})} 
    e^{-i\int\mathrm{d}^4x\,\delta H_\text{higher}}
    \no\\
    & \quad\times
    \prod_{k\leq k_c(T)}
    \delta(\phi^\text{IR}_\Delta(T,\bm k))
    \delta(\phi^\text{IR}_c(t_k,\bm k)-\phi_\text{false}(2\pi)^3\delta(\bm k))
    \delta(\Pi^\text{IR}_c(t_k-0^+,\bm k))
   \,,
    \label{IRgene5-1}
\end{align}
where, $\displaystyle\widetilde{\mathcal{D}}\Pi_c^\text{IR}\coloneqq\prod_{k\leq k_c(T)}\mathrm{d}\Pi^\text{IR}_c(t_k-0^+,\bm k)$. Below, we simply omit this integration measure just for simplifying the notation. We also decomposed $S_{H}^\text{IR}$ as $S_H^\text{IR}=S_{H,(d)}^\text{IR}+S_{H,(s)}^\text{IR}$ up to boundary terms which vanish thanks to the trick \eqref{trick2} as 
\begin{align}
    S^\text{IR}_{H,(d)}
    &=
    \int\mathrm{d}^4x\,
    \left[
    \Pi^\text{IR}_\Delta
    \left(
        \dot \phi^\text{IR}_c-a^{-3}\Pi^\text{IR}_c 
    \right)
    -\phi^\text{IR}_\Delta
    \left(
        \dot\Pi^\text{IR}_c-a(t)\nabla^2\phi^\text{IR}_c
        +a^3(t)\,V'(\phi^\text{IR}_c)
    \right)
    \right]
    \,,\label{action_d}
    \\
    S^\text{IR}_{H,(s)}
    &=
    -\int\mathrm{d}^4x\,a^3(t)\,
    \left[
        V(\phi^\text{IR}_c+(\phi^\text{IR}_\Delta/2))
        -
        V(\phi^\text{IR}_c-(\phi^\text{IR}_\Delta/2))
        -
        V'(\phi^\text{IR}_c)\phi^\text{IR}_\Delta
    \right]\,.
\end{align}
Here, the space-time argument of IR variables are omitted. We introduced the IR variables and their time derivatives as 
\begin{align}
    &X^\text{IR}(t,\bm x)
    \coloneqq
    \int\frac{\mathrm{d}^3k}{(2\pi)^3}\,
    \theta(k_c(t)-k)\,
    X^\text{IR}(t,\bm k)\, e^{i\bm k.\bm x}
    \,,
    \\
&    \dot X^\text{IR}(t,\bm x)
    \coloneqq
    \int\frac{\mathrm{d}^3k}{(2\pi)^3}\,
    \theta(k_c(t)-k)\,
    \dot X^\text{IR}(t,\bm k)\, e^{i\bm k.\bm x}
    \,,\label{real_timederiv}
\end{align}
where $X^\text{IR}$ represents the IR variables: $X^\text{IR}=(\phi^\text{IR}_c,\phi^\text{IR}_\Delta,\Pi^\text{IR}_c,\Pi^\text{IR}_\Delta)$. Precisely speaking, for $X^\text{IR}=\Pi^\text{IR}_c$, we should replace $k_c(t)$ by $k_c(t+0^+)$ so that we have $k_c(t)=k$ at $t=t_k-0^+$. 
One may concern about the subtlety of the time derivatives $\dot\phi^\text{IR}_c(t,\bm x)$ and $\dot\Pi^\text{IR}_c(t,\bm x)$ in \eqref{action_d}: We in general have $\dot X^\text{IR}(t,\bm x)\neq \frac{\partial}{\partial t}(X^\text{IR}(t,\bm x))$, where the LHS is defined by \eqref{real_timederiv}, due to the presence of time-dependent step function. However, there is no subtlety thanks to the initial conditions for all IR modes $k\leq k_c(T)$ of the form
\begin{align}
   \phi^\text{IR}_c(t_k,\bm k)
   = \phi_\text{false}(2\pi)^3\delta(\bm k)\,,
   \qquad
   \Pi^\text{IR}_c(t_k-0^+,\bm k) = 0
   \,,\label{initial}
\end{align}
which are imposed by the delta functions in the final line in \eqref{IRgene5-1}. The corresponding conditions in the real space for all $\bm x$ are 
\begin{subequations}
\label{initial_real}
\begin{align}
    &\dot X^\text{IR}(t,\bm x)
    = \partial_t (X^\text{IR}(t,\bm x)) 
    \quad \text{for}
    \quad X^\text{IR}=(\phi^\text{IR}_c,\,\Pi^\text{IR}_c)
    \,,\\
    &\phi^\text{IR}_c(-\infty,\bm x)
    =\phi_\text{false}
    \,,\quad
    \Pi^\text{IR}_c(-\infty,\bm x)
    = 0\,.
    \label{initial_realb}
\end{align}
\end{subequations}
To implement these conditions in the path integral, we define $\delta\left[\mathcal{C}_\text{ini}\right]$ by \eqref{pathint_bc_ini} with
\begin{equation}
    \delta\left[\mathcal{C}_\text{detail}\right]
    \coloneqq
    \prod_{-\infty\leq t\leq T}\prod_{X^\text{IR}=(\phi^\text{IR}_c,\Pi^\text{IR}_c)}
    \prod_{\bm x}
    \delta
    \left(
        \dot X^\text{IR}(t,\bm x) - \partial_t (X^\text{IR}(t,\bm x))
    \right)
    \,.
\end{equation}
We find from eqs.~\eqref{tunnel_IRgene} and \eqref{IRgene5-1} that the transition probability can be written in terms of path integral as
\begin{align}
    p(\{\phi^\text{IR}(T,\bm{x})\}_{T,\bm x\in\mathcal{D}})
    =
    &\int\mathcal{D}(\phi^\text{IR}_c,\phi^\text{IR}_\Delta,\Pi^\text{IR}_c,\Pi^\text{IR}_\Delta)
    \,
    \,e^{i\mathcal{S}}e^{i\Gamma_{k=0}}
    \delta\left[\mathcal{C}_\text{fin}\right]
    \delta\left[\mathcal{C}_\text{ini}\right]
   \,,\label{tunnel_IRgene2}
\end{align}
where $\delta\left[\mathcal{C}_\text{fin}\right]$ is defined in \eqref{pathint_bc_fin}, and the term $\mathcal S$ in the exponent is given by 

\begin{equation}
    \mathcal S 
    = (S_{H,(d)}^\text{IR}+S_{H,(s)}^\text{IR}+\Gamma_\text{LO}^{k>0})-\int\mathrm{d}^4x\,\delta H_\text{higher}
\end{equation}
which agrees with the definition of $\mathcal{S}$ given in \eqref{eq:effective_action}. When fluctuations of zero mode are negligible, we can set $\Gamma_{k=0}=0$. This is the setup adopted in the main text. Then, \eqref{tunnel_IRgene2} coincides with \eqref{tunnel_IRgene_main}.

\subsection{Stochastic interpretation of the tunneling}\label{sec:interpretation}
It is useful to see how the stochastic equations emerge as a consequence of performing the path integral over IR variables non-perturbatively. This leads to the stochastic interpretation of the tunneling probability. A final result of this section is \eqref{stochastic_tunnel}.

For simplicity, let us ignore higher-order corrections $\delta H_\text{higher}$ for a while. We start with introducing the auxiliary fields $(\xi_\phi,\xi_\Pi)$ as 
\begin{align}
    e^{i\left(\Gamma^{k>0}_\text{LO}+\Gamma_{k=0}+S^\text{IR}_{H,(s)}\right)}
    =
    \int\mathcal{D}(\xi_\phi,\xi_\Pi)
    \,P[\xi_\phi,\xi_\Pi]
    e^{i\int\mathrm{d}^4x
    \left(
        \xi_\Pi(x)\phi^\text{IR}_\Delta(x)
        -
        \xi_\phi(x)\Pi^\text{IR}_\Delta(x)
    \right)
    }
    \,.\label{noise_definition}
\end{align}
We substitute \eqref{noise_definition} into \eqref{IRgene5-1}. Then the exponent becomes linear in $\phi^\text{IR}_\Delta(t,\bm k)$ or $\Pi^\text{IR}_\Delta(t,\bm k)$ with $t_0\leq t<T$: 
\begin{align}
    Z[J^\text{IR}(T)]
    &\simeq
    \int\mathcal{D}(\phi^\text{IR}_c,\phi^\text{IR}_\Delta,\Pi^\text{IR}_c,\Pi^\text{IR}_\Delta)
    \int\widetilde{\mathcal{D}}\Pi^\text{IR}_c
    \,
    e^{i \bm{J}^\text{IR}\cdot\bm\phi^\text{IR}_c}
    \int\mathcal{D}(\xi_\phi,\xi_\Pi)\,
    P[\xi_\phi,\xi_\Pi]
     \no\\
    &\times
    \exp
    \left\{
    i\int^T_{-\infty}\mathrm{d}t\int\frac{\mathrm{d}^3k}{(2\pi)^3}\,
    \left[
        \Pi^\text{IR}_\Delta(t,-\bm k)\mathcal{L}_\phi(t,\bm k)
        -
        \phi^\text{IR}_\Delta(t,-\bm k)\mathcal{L}_\Pi(t,\bm k)
    \right]
    \Theta(k_c(t)-k)
    \right\}
    \no\\
    &\times
    \prod_{k\leq k_c(T)}
    \delta(\phi^\text{IR}_\Delta(T,\bm k))
    \delta(\phi^\text{IR}_c(t_k,\bm k)-\phi_\text{false}(2\pi)^3\delta(\bm k))
    \delta(\Pi^\text{IR}_c(t_k-0^+,\bm k))
   \,,
    \label{IRgene6}
\end{align}
where 
\begin{subequations}
\label{Langevin1}
\begin{align}
    &\mathcal{L}_\phi(t,\bm k) 
    \coloneqq
    \dot\phi^\text{IR}_c(t,\bm k)-a^{-3}(t)\Pi^\text{IR}_c(t,\bm k)-\xi_\phi(t,\bm k)\,,
    \\
    &\mathcal{L}_\pi(t,\bm k)
    \coloneqq
    \dot\Pi^\text{IR}_c(t,\bm k)+a(t)k^2\phi^\text{IR}_c(t,\bm k)
        +a^3(t)\,V'(\phi^\text{IR}_c)(t,\bm k)-\xi_\Pi(t,\bm k)
    \,.
\end{align}
\end{subequations}
Consequently, we can perform the path integral $\int\mathcal{D}(\phi^\text{IR}_\Delta,\Pi^\text{IR}_\Delta)$, leading to
\begin{align}
    Z[J^\text{IR}(T)]
    \simeq
    \prod_{k\leq k_c(T)}
    \left.
    \int\mathrm{d}\phi^\text{IR}_c(T,\bm k)
    \,e^{i \bm{J}^\text{IR}\cdot\bm\phi^\text{IR}_c}
    \int\mathcal{D}(\xi_\phi,\xi_\Pi)\,
    P[\xi_\phi,\xi_\Pi]
    \right|_\text{$\mathcal{L}_\phi=\mathcal{L}_\Pi=0$ \& \eqref{initial}}
    \,.\label{IRgene7}
\end{align} 
Now we regard auxiliary fields $(\xi_\phi,\xi_\Pi)$ as noise variables whose correlations are given by 
\begin{align}
    \langle\xi_\alpha(t,\bm k) \xi_\beta(t',\bm k')\rangle
    &\coloneqq
    \int\mathcal{D}(\xi_\phi,\xi_\Pi)
    \,P[\xi_\phi,\xi_\Pi]\,
    \xi_\alpha(t,\bm k)\xi_\beta(t',\bm k')
    \no\\
    &
    =
    \left.
    \left[
        \frac{i\delta}{\delta X_\alpha(t,\bm k)}
        \frac{i\delta}{\delta X_\beta(t',\bm k')}
    \right]
    e^{i\left(\Gamma^{k>0}_\text{LO}+\Gamma_{k=0}+S^\text{IR}_{H,(s)}\right)}
    \right|_{\phi^\text{IR}_\Delta=\Pi^\text{IR}_\Delta=0}
    \,,\label{noise1}
\end{align}
where $\alpha,\beta = (\phi,\Pi)$ and $(X_\phi,X_\Pi)=(\Pi^\text{IR}_\Delta, -\phi^\text{IR}_\Delta)$ following the notation in \eqref{infl3}. We also have the corresponding expression in the momentum space. If we ignore $S_{H,(s)}$ and $i\Gamma_{k=0}$, we have 
\begin{equation}
    \langle\xi_\alpha(t,\bm k) \xi_\beta(t',\bm k')\rangle
    =
    (2\pi)^3\delta\left(\bm k+\bm k'\right) 
    \text{Re}\left[g^{\alpha\beta}(t)\right]\delta(t-t')\delta(t-t_k) 
    \,.\label{noise2_mom}
\end{equation}
Eq.\eqref{IRgene7} shows that the effective dynamics of IR fields are described by the set of Langevin equations $\mathcal{L}_\phi=\mathcal{L}_\Pi=0$ with the initial condition \eqref{initial} and the noise correlations \eqref{noise2_mom}.

In the real space, the dynamics is described by the set of Langevin equations,
\begin{subequations}
\label{langevin_real1}
 \begin{align}
    &\dot\phi^\text{IR}_c(x)=a^{-3}(t)\Pi^\text{IR}_c(x)+\xi_\phi(x)
    \,,\\
    &\dot\Pi^\text{IR}_c(x)=a(t)\nabla^2\phi^\text{IR}_c(x)-a^3V'(\phi^\text{IR}_c(x))+\xi_\Pi(x)
    \,,
\end{align}
\end{subequations}
with the initial conditions \eqref{initial_real} and the noise correlations \begin{equation}
    \langle\xi_\alpha(x) \xi_\beta(y)\rangle
    = G^{\alpha\beta}(x,y)
    \,.\label{noise2}
\end{equation}
Eq.\eqref{IRgene7} takes the following form in the real space:
\begin{align}
    Z[J^\text{IR}(T)]
    \simeq
    \prod_{\bm x}
    \left.
    \int\mathrm{d}\phi^\text{IR}_c(T,\bm x)
    \,e^{i \bm{J}^\text{IR}\cdot\bm\phi^\text{IR}_c}
    \int\mathcal{D}(\xi_\phi,\xi_\Pi)\,
    P[\xi_\phi,\xi_\Pi]
    \right|_\text{\eqref{initial_real} \& \eqref{langevin_real1}}
    \,.\label{IRgene7_real}
\end{align} 
Substituting \eqref{IRgene7_real} into \eqref{tunnel_IRgene}, we have
 \begin{align}
    p(\{\phi^\text{IR}(T,\bm{x})\}_{T,\bm x\in\mathcal{D}})
    \simeq 
    &\prod_{\bm y}
    \int\mathrm{d}\phi^\text{IR}_c(T,\bm y)
    \,\prod_{\bm x\in\mathcal{D}}\delta(\phi^\text{IR}_c(T,\bm x)-\phi^\text{IR}(T,\bm x))
    \no\\
    &\quad\times
    \left.
    \int\mathcal{D}(\xi_\phi,\xi_\Pi)\,
    P[\xi_\phi,\xi_\Pi]
    \right|_\text{\eqref{initial_real} \& \eqref{langevin_real1}}
    \,.\label{stochastic_tunnel}
 \end{align}
This expression provides a stochastic interpretation of the transition probability; 
the dynamics of $\phi^\text{IR}_c$ is described by the set of Langevin equations and the transition probability is simply understood as the probability to realize the field configuration $\phi^\text{IR}_c(T,\bm x)=\phi^\text{IR}(T,\bm x)$ in the domain $\bm x\in\mathcal{D}$ at the final time $T$ starting from the initial configuration $\phi^\text{IR}_c(-\infty,\bm x)=\phi_\text{false}$. In particular, the set of Langevin equations is given by \eqref{eq:LangevinFull} with the noise correlations \eqref{symmetric_noise} when we ignore the terms $\delta H_\text{higher}$, $S_{H,(s)}^\text{IR}$, and $\Gamma_{k=0}$. Note that such terms simply correct the Langevin equations\footnote{For instance, leading-order corrections to the stochastic dynamics at the super-horizon scales are calculated in \cite{Tokuda:2017fdh}.} and do not invalidate the above stochastic interpretation.

\section{Coleman - de Luccia tunneling in Euclidean method}\label{sec:Euclidean_CDL_appendix}

In this appendix, we provide the CDL action in de Sitter background with Euclidean method.

Generally, the ways of Euclideanization for de Sitter spacetime depend on the coordinates. In this paper, we discuss the field theory on the metric \eqref{eq:Metric}, and then we choose the following coordinate again \footnote{The analytic continuations of this coordinate is examined in \cite{Rubakov:1999ir}}.
\begin{align}
    \dm s^2=\frac{1}{H^2\eta^2}(-\dm\eta^2+\dm\bm{x}^2).
\end{align}
By taking the wick rotation as $\eta\to -i\tau$, the coordinate becomes
\begin{align}
    -\dm s^2=\frac{1}{H^2\tau^2}(\dm\tau^2+\dm\bm{x}^2).
\end{align}
This is the Euclidean AdS space discussed in section~\ref{sec:CDL} and we can take the global coordinate;
\begin{align}
    -\dm s^2=H^{-2}(\dm\rho^2 + \sinh^2\rho \dm\Omega^2).
\end{align}
We also impose that the field $\phi$ only depends on $\rho$. Therefore, the Euclidean action $I^E$ and the equation of motion becomes
\begin{align}
    I^E=2\pi^2\int_0^\infty \dm\rho \sinh^3\rho \bigg[\frac{1}{2H^2}(\partial_\rho\phi)^2 + \frac{V(\phi)}{H^4}\bigg], \label{eq:Euclidean_Action}\\
    \frac{\dm^{2}\phi}{\dm\rho^{2}} + \frac{3}{\tanh\rho}\frac{\dm\phi}{\dm\rho} = \frac{V'(\phi)}{H^{2}}.
\end{align}
The second equation is the same as \eqref{eq:HamiltonEqO4} and we can obtain the same bounce solutions discussed in section~\ref{sec:CDL}.
For numerical calculations, we consider the potential \eqref{eq:potential} and fitting function \eqref{eq:Fitting}. Also we take the following non-dimensionalizations;
\begin{align}
 \phi &= \alpha\tilde{\phi}, \\
 V(\phi)&=g^2\alpha^4 \tilde{V}(\tilde{\phi}).
\end{align}
Therefore, the Euclidean action \eqref{eq:Euclidean_Action} becomes
\begin{align}
    I^E&=2\pi^2 \frac{\alpha^2}{H^2}\int_0^\infty \dm\rho \sinh^3\rho \bigg[\frac{1}{2}(\partial_\rho\tilde{\phi})^2 + \frac{g^2\alpha^2}{H^2}\tilde{V}(\tilde{\phi})\bigg] \notag\\
    &:=2\pi^2 \frac{\alpha^2}{H^2} \tilde{I}^E.
    \label{eq:tilde_Euclid}
\end{align}
The ratio of the Coleman-de Luccia Euclidean action $I^{E}_{\text{bounce}}$ to the Hawking-Moss action $I_{\text{HM}}$ is given as
\begin{align}
    \gamma_E = \frac{I^E_{\text{bounce}}}{I_\text{HM}}=\frac{H^2}{g^2\alpha^2}\frac{9\tilde{I}^E_{\text{bounce}}}{(1-\beta)^3(\beta+3)}. \label{eq:ratio_E}
\end{align}
%

 \bibliographystyle{JHEP}
 \bibliography{stochastic.bib}

\providecommand{\href}[2]{#2}\begingroup\raggedright\begin{thebibliography}{10}

\bibitem{Coleman:1977py}
S.R.~Coleman, \emph{{The Fate of the False Vacuum. 1. Semiclassical Theory}},
  \href{https://doi.org/10.1103/PhysRevD.16.1248}{\emph{Phys. Rev. D}
  {\bfseries 15} (1977) 2929}.

\bibitem{Callan:1977pt}
C.G.~Callan, Jr. and S.R.~Coleman, \emph{{The Fate of the False Vacuum. 2.
  First Quantum Corrections}},
  \href{https://doi.org/10.1103/PhysRevD.16.1762}{\emph{Phys. Rev. D}
  {\bfseries 16} (1977) 1762}.

\bibitem{Coleman:1980aw}
S.R.~Coleman and F.~De~Luccia, \emph{{Gravitational Effects on and of Vacuum
  Decay}}, \href{https://doi.org/10.1103/PhysRevD.21.3305}{\emph{Phys. Rev. D}
  {\bfseries 21} (1980) 3305}.

\bibitem{Hawking:1981fz}
S.W.~Hawking and I.G.~Moss, \emph{{Supercooled Phase Transitions in the Very
  Early Universe}},
  \href{https://doi.org/10.1016/0370-2693(82)90946-7}{\emph{Phys. Lett. B}
  {\bfseries 110} (1982) 35}.

\bibitem{Rubakov:1999ir}
V.A.~Rubakov and S.M.~Sibiryakov, \emph{{False vacuum decay in de Sitter
  space-time}}, \href{https://doi.org/10.1007/BF02557243}{\emph{Theor. Math.
  Phys.} {\bfseries 120} (1999) 1194}
  [\href{https://arxiv.org/abs/gr-qc/9905093}{{\ttfamily gr-qc/9905093}}].

\bibitem{Brown:2007sd}
A.R.~Brown and E.J.~Weinberg, \emph{{Thermal derivation of the Coleman-De
  Luccia tunneling prescription}},
  \href{https://doi.org/10.1103/PhysRevD.76.064003}{\emph{Phys. Rev. D}
  {\bfseries 76} (2007) 064003}
  [\href{https://arxiv.org/abs/0706.1573}{{\ttfamily 0706.1573}}].

\bibitem{Starobinsky:1986fx}
A.A.~Starobinsky, \emph{{STOCHASTIC DE SITTER (INFLATIONARY) STAGE IN THE EARLY
  UNIVERSE}}, \href{https://doi.org/10.1007/3-540-16452-9_6}{\emph{Lect. Notes
  Phys.} {\bfseries 246} (1986) 107}.

\bibitem{Starobinsky:1994bd}
A.A.~Starobinsky and J.~Yokoyama, \emph{{Equilibrium state of a selfinteracting
  scalar field in the De Sitter background}},
  \href{https://doi.org/10.1103/PhysRevD.50.6357}{\emph{Phys. Rev. D}
  {\bfseries 50} (1994) 6357}
  [\href{https://arxiv.org/abs/astro-ph/9407016}{{\ttfamily
  astro-ph/9407016}}].

\bibitem{Goncharov:1987ir}
A.S.~Goncharov, A.D.~Linde and V.F.~Mukhanov, \emph{{The Global Structure of
  the Inflationary Universe}},
  \href{https://doi.org/10.1142/S0217751X87000211}{\emph{Int. J. Mod. Phys. A}
  {\bfseries 2} (1987) 561}.

\bibitem{Linde:1991sk}
A.D.~Linde, \emph{{Hard art of the universe creation (stochastic approach to
  tunneling and baby universe formation)}},
  \href{https://doi.org/10.1016/0550-3213(92)90326-7}{\emph{Nucl. Phys. B}
  {\bfseries 372} (1992) 421}
  [\href{https://arxiv.org/abs/hep-th/9110037}{{\ttfamily hep-th/9110037}}].

\bibitem{Tolley:2008qv}
A.J.~Tolley and M.~Wyman, \emph{{Stochastic tunneling in DBI inflation}},
  \href{https://doi.org/10.1088/1475-7516/2009/10/006}{\emph{JCAP} {\bfseries
  10} (2009) 006} [\href{https://arxiv.org/abs/0809.1100}{{\ttfamily
  0809.1100}}].

\bibitem{Noorbala:2018zlv}
M.~Noorbala, V.~Vennin, H.~Assadullahi, H.~Firouzjahi and D.~Wands,
  \emph{{Tunneling in Stochastic Inflation}},
  \href{https://doi.org/10.1088/1475-7516/2018/09/032}{\emph{JCAP} {\bfseries
  09} (2018) 032} [\href{https://arxiv.org/abs/1806.09634}{{\ttfamily
  1806.09634}}].

\bibitem{Hashiba:2020rsi}
S.~Hashiba, Y.~Yamada and J.~Yokoyama, \emph{{Particle production induced by
  vacuum decay in real time dynamics}},
  \href{https://doi.org/10.1103/PhysRevD.103.045006}{\emph{Phys. Rev. D}
  {\bfseries 103} (2021) 045006}
  [\href{https://arxiv.org/abs/2006.10986}{{\ttfamily 2006.10986}}].

\bibitem{Camargo-Molina:2022ord}
J.E.~Camargo-Molina and A.~Rajantie, \emph{{Phase transitions in de Sitter: The
  stochastic formalism}},  \href{https://arxiv.org/abs/2204.02875}{{\ttfamily
  2204.02875}}.

\bibitem{Camargo-Molina:2022paw}
J.E.~Camargo-Molina, M.C.~Gonz\'alez and A.~Rajantie, \emph{{Phase Transitions
  in de Sitter: Quantum Corrections}},
  \href{https://arxiv.org/abs/2204.03480}{{\ttfamily 2204.03480}}.

\bibitem{Martin:1973zz}
P.C.~Martin, E.D.~Siggia and H.A.~Rose, \emph{{Statistical Dynamics of
  Classical Systems}},
  \href{https://doi.org/10.1103/PhysRevA.8.423}{\emph{Phys. Rev. A} {\bfseries
  8} (1973) 423}.

\bibitem{DeDominicis:1975gjb}
C.~De~Dominicis, E.~Brezin and J.~Zinn-Justin, \emph{{Field Theoretic
  Techniques and Critical Dynamics. 1. Ginzburg-Landau Stochastic Models
  Without Energy Conservation}},
  \href{https://doi.org/10.1103/PhysRevB.12.4945}{\emph{Phys. Rev. B}
  {\bfseries 12} (1975) 4945}.

\bibitem{Janssen:1976aa}
H.-K.~Janssen, \emph{On a lagrangean for classical field dynamics and
  renormalization group calculations of dynamical critical properties},
  \href{https://doi.org/10.1007/BF01316547}{\emph{Zeitschrift f{\"u}r Physik B
  Condensed Matter} {\bfseries 23} (1976) 377}.

\bibitem{DeDominicis:1977fw}
C.~De~Dominicis and L.~Peliti, \emph{{Field Theory Renormalization and Critical
  Dynamics Above t(c): Helium, Antiferromagnets and Liquid Gas Systems}},
  \href{https://doi.org/10.1103/PhysRevB.18.353}{\emph{Phys. Rev. B} {\bfseries
  18} (1978) 353}.

\bibitem{PhysRevE.70.041106}
V.~Elgart and A.~Kamenev, \emph{Rare event statistics in reaction-diffusion
  systems}, \href{https://doi.org/10.1103/PhysRevE.70.041106}{\emph{Phys. Rev.
  E} {\bfseries 70} (2004) 041106}.

\bibitem{altland_simons_2010}
A.~Altland and B.D.~Simons, \emph{Condensed Matter Field Theory}, Cambridge
  University Press, 2~ed. (2010),
  \href{https://doi.org/10.1017/CBO9780511789984}{10.1017/CBO9780511789984}.

\bibitem{Tokuda:2017fdh}
J.~Tokuda and T.~Tanaka, \emph{{Statistical nature of infrared dynamics on de
  Sitter background}},
  \href{https://doi.org/10.1088/1475-7516/2018/02/014}{\emph{JCAP} {\bfseries
  02} (2018) 014} [\href{https://arxiv.org/abs/1708.01734}{{\ttfamily
  1708.01734}}].

\bibitem{Tokuda:2018eqs}
J.~Tokuda and T.~Tanaka, \emph{{Can all the infrared secular growth really be
  understood as increase of classical statistical variance?}},
  \href{https://doi.org/10.1088/1475-7516/2018/11/022}{\emph{JCAP} {\bfseries
  11} (2018) 022} [\href{https://arxiv.org/abs/1806.03262}{{\ttfamily
  1806.03262}}].

\bibitem{Morikawa:1989xz}
M.~Morikawa, \emph{{Dissipation and Fluctuation of Quantum Fields in Expanding
  Universes}}, \href{https://doi.org/10.1103/PhysRevD.42.1027}{\emph{Phys. Rev.
  D} {\bfseries 42} (1990) 1027}.

\bibitem{Feynman:1963fq}
R.P.~Feynman and F.L.~Vernon, Jr., \emph{{The Theory of a general quantum
  system interacting with a linear dissipative system}},
  \href{https://doi.org/10.1016/0003-4916(63)90068-X}{\emph{Annals Phys.}
  {\bfseries 24} (1963) 118}.

\bibitem{Weinberg:2012pjx}
E.J.~Weinberg, \emph{Classical Solutions in Quantum Field Theory: Solitons and
  Instantons in High Energy Physics}, Cambridge Monographs on Mathematical
  Physics, Cambridge University Press (2012),
  \href{https://doi.org/10.1017/CBO9781139017787}{10.1017/CBO9781139017787}.

\bibitem{Weinberg:2006pc}
E.J.~Weinberg, \emph{{Hawking-Moss bounces and vacuum decay rates}},
  \href{https://doi.org/10.1103/PhysRevLett.98.251303}{\emph{Phys. Rev. Lett.}
  {\bfseries 98} (2007) 251303}
  [\href{https://arxiv.org/abs/hep-th/0612146}{{\ttfamily hep-th/0612146}}].

\bibitem{Jensen:1983ac}
L.G.~Jensen and P.J.~Steinhardt, \emph{{Bubble Nucleation and the
  {Coleman-Weinberg} Model}},
  \href{https://doi.org/10.1016/0550-3213(84)90021-X}{\emph{Nucl. Phys. B}
  {\bfseries 237} (1984) 176}.

\bibitem{Jensen:1988zx}
L.G.~Jensen and P.J.~Steinhardt, \emph{{Bubble Nucleation for Flat Potential
  Barriers}}, \href{https://doi.org/10.1016/0550-3213(89)90539-7}{\emph{Nucl.
  Phys. B} {\bfseries 317} (1989) 693}.

\bibitem{Hackworth:2004xb}
J.C.~Hackworth and E.J.~Weinberg, \emph{{Oscillating bounce solutions and
  vacuum tunneling in de Sitter spacetime}},
  \href{https://doi.org/10.1103/PhysRevD.71.044014}{\emph{Phys. Rev. D}
  {\bfseries 71} (2005) 044014}
  [\href{https://arxiv.org/abs/hep-th/0410142}{{\ttfamily hep-th/0410142}}].

\bibitem{Batra:2006rz}
P.~Batra and M.~Kleban, \emph{{Transitions Between de Sitter Minima}},
  \href{https://doi.org/10.1103/PhysRevD.76.103510}{\emph{Phys. Rev. D}
  {\bfseries 76} (2007) 103510}
  [\href{https://arxiv.org/abs/hep-th/0612083}{{\ttfamily hep-th/0612083}}].

\bibitem{Braden:2018tky}
J.~Braden, M.C.~Johnson, H.V.~Peiris, A.~Pontzen and S.~Weinfurtner, \emph{{New
  Semiclassical Picture of Vacuum Decay}},
  \href{https://doi.org/10.1103/PhysRevLett.123.031601}{\emph{Phys. Rev. Lett.}
  {\bfseries 123} (2019) 031601}
  [\href{https://arxiv.org/abs/1806.06069}{{\ttfamily 1806.06069}}].

\bibitem{Blanco-Pillado:2019xny}
J.J.~Blanco-Pillado, H.~Deng and A.~Vilenkin, \emph{{Flyover vacuum decay}},
  \href{https://doi.org/10.1088/1475-7516/2019/12/001}{\emph{JCAP} {\bfseries
  12} (2019) 001} [\href{https://arxiv.org/abs/1906.09657}{{\ttfamily
  1906.09657}}].

\bibitem{Hertzberg:2020tqa}
M.P.~Hertzberg, F.~Rompineve and N.~Shah, \emph{{Quantitative Analysis of the
  Stochastic Approach to Quantum Tunneling}},
  \href{https://doi.org/10.1103/PhysRevD.102.076003}{\emph{Phys. Rev. D}
  {\bfseries 102} (2020) 076003}
  [\href{https://arxiv.org/abs/2009.00017}{{\ttfamily 2009.00017}}].

\bibitem{Tranberg:2022noe}
A.~Tranberg and G.~Ungersb\"ack, \emph{{Bubble nucleation and quantum initial
  conditions in classical statistical simulations}},
  \href{https://doi.org/10.1007/JHEP09(2022)206}{\emph{JHEP} {\bfseries 09}
  (2022) 206} [\href{https://arxiv.org/abs/2206.08691}{{\ttfamily
  2206.08691}}].

\bibitem{Hertzberg:2019wgx}
M.P.~Hertzberg and M.~Yamada, \emph{{Vacuum Decay in Real Time and Imaginary
  Time Formalisms}},
  \href{https://doi.org/10.1103/PhysRevD.100.016011}{\emph{Phys. Rev. D}
  {\bfseries 100} (2019) 016011}
  [\href{https://arxiv.org/abs/1904.08565}{{\ttfamily 1904.08565}}].

\bibitem{Kristiano:2018oyv}
J.~Kristiano, R.D.~Lambaga and H.S.~Ramadhan, \emph{{Coleman-de Luccia
  Tunneling Wave Function}},
  \href{https://doi.org/10.1016/j.physletb.2019.07.040}{\emph{Phys. Lett. B}
  {\bfseries 796} (2019) 225}
  [\href{https://arxiv.org/abs/1808.10110}{{\ttfamily 1808.10110}}].

\bibitem{Cespedes:2020xpn}
S.~Cespedes, S.P.~de~Alwis, F.~Muia and F.~Quevedo, \emph{{Lorentzian vacuum
  transitions: Open or closed universes?}},
  \href{https://doi.org/10.1103/PhysRevD.104.026013}{\emph{Phys. Rev. D}
  {\bfseries 104} (2021) 026013}
  [\href{https://arxiv.org/abs/2011.13936}{{\ttfamily 2011.13936}}].

\bibitem{Maniccia:2022iqa}
G.~Maniccia, M.~De~Angelis and G.~Montani, \emph{{WKB Approaches to Restore
  Time in Quantum Cosmology: Predictions and Shortcomings}},
  \href{https://doi.org/10.3390/universe8110556}{\emph{Universe} {\bfseries 8}
  (2022) 556} [\href{https://arxiv.org/abs/2209.04403}{{\ttfamily
  2209.04403}}].

\bibitem{Espinosa:2018voj}
J.R.~Espinosa, \emph{{Fresh look at the calculation of tunneling actions
  including gravitational effects}},
  \href{https://doi.org/10.1103/PhysRevD.100.104007}{\emph{Phys. Rev. D}
  {\bfseries 100} (2019) 104007}
  [\href{https://arxiv.org/abs/1808.00420}{{\ttfamily 1808.00420}}].

\bibitem{Gregory:2020cvy}
R.~Gregory, I.G.~Moss and N.~Oshita, \emph{{Black Holes, Oscillating
  Instantons, and the Hawking-Moss transition}},
  \href{https://doi.org/10.1007/JHEP07(2020)024}{\emph{JHEP} {\bfseries 07}
  (2020) 024} [\href{https://arxiv.org/abs/2003.04927}{{\ttfamily
  2003.04927}}].

\bibitem{Gregory:2020hia}
R.~Gregory, I.G.~Moss, N.~Oshita and S.~Patrick, \emph{{Hawking-Moss transition
  with a black hole seed}},
  \href{https://doi.org/10.1007/JHEP09(2020)135}{\emph{JHEP} {\bfseries 09}
  (2020) 135} [\href{https://arxiv.org/abs/2007.11428}{{\ttfamily
  2007.11428}}].

\bibitem{Freese:2004vs}
K.~Freese and D.~Spolyar, \emph{{Chain inflation: 'Bubble bubble toil and
  trouble'}}, \href{https://doi.org/10.1088/1475-7516/2005/07/007}{\emph{JCAP}
  {\bfseries 07} (2005) 007}
  [\href{https://arxiv.org/abs/hep-ph/0412145}{{\ttfamily hep-ph/0412145}}].

\bibitem{Berera:1995ie}
A.~Berera, \emph{{Warm inflation}},
  \href{https://doi.org/10.1103/PhysRevLett.75.3218}{\emph{Phys. Rev. Lett.}
  {\bfseries 75} (1995) 3218}
  [\href{https://arxiv.org/abs/astro-ph/9509049}{{\ttfamily
  astro-ph/9509049}}].

\end{thebibliography}\endgroup

\end{document}